\newcommand{\preprintDate}{28 November 2022}

\documentclass{jfm-arxiv}

\usepackage[T1]{fontenc}
\usepackage[utf8]{inputenc}
\usepackage[UKenglish]{babel}
\usepackage[style=british]{csquotes}
\usepackage{microtype}

\usepackage{graphicx}
\usepackage{newtxtext}
\usepackage{newtxmath}

\usepackage[breaklinks]{hyperref}
\hypersetup{
    colorlinks = true,
    urlcolor   = blue,
    citecolor  = black,
}

\usepackage[backend=biber,natbib=true,style=authoryear-comp,sorting=nyvt,sortlocale=en-GB,maxbibnames=999,maxcitenames=2,mincitenames=1,date=year,labeldate=year,alldates=year]{biblatex} 
\bibliography{srdmd}
\renewbibmacro{in:}{}

\usepackage{overpic}
\usepackage{mathtools}
\usepackage{ifthen}
\usepackage{gensymb}
\usepackage[normalem]{ulem} 
\usepackage{enumitem}

\graphicspath{{fig/}}

\makeatletter
\g@addto@macro\@floatboxreset{\centering}
\makeatother
\newcommand{\secref}[1]{\S\ref{sec/#1}}
\newcommand{\figref}[1]{figure~\ref{fig/#1}}
\newcommand{\tabref}[1]{table~\ref{tab/#1}}
\newcommand{\Figref}[1]{Figure~\ref{fig/#1}}

\newcommand{\fig}[3][1]{
    \begin{figure}
        \includegraphics[width=#1\textwidth]{fig/#2}
        \caption{#3 \label{fig/#2}}
    \end{figure}
    }

\newcommand{\tab}[5]{ 
    \begin{table}     
        \caption{
            #2 \label{tab/#1}
        }    
        \begin{tabular}{#3} 
        \hline 
            #4 \\
        \hline
         #5 
        \end{tabular}
    \end{table}
}

\newcommand{\eq}[2]{
    \begin{equation}
        #2 \label{#1}
    \end{equation}
}

\newcommand{\eqarr}[1]{
    \begin{eqnarray}
        #1
    \end{eqnarray}
}

\newcommand{\bracket}[1]{
    \left[ #1 \right]
}

\DeclarePairedDelimiter\autoparentheses{(}{)}
\newcommand{\pa}[1]{\autoparentheses*{#1}}

\newcommand{\de}{\ensuremath{\delta}}
\newcommand{\De}{\ensuremath{\Delta}}

\newcommand{\la}{\ensuremath{\lambda}}
\newcommand{\La}{\ensuremath{\Lambda}}

\newcommand{\Si}{\ensuremath{\Sigma}}
\newcommand{\ta}{\ensuremath{\tau}}

\newcommand{\om}{\ensuremath{\omega}}

\newcommand{\ps}{\ensuremath{\psi}}

\newcommand\flow[2]{{\Phi^{#1}(#2)}}

\newcommand{\identity}{\ensuremath{I}}

\newcommand{\TW}{\ensuremath{\mathrm{TW}}}
\newcommand{\RPO}{\ensuremath{\mathrm{RPO}}}
\newcommand{\ts}{\ensuremath{t_s}} 
\newcommand{\sldt}{\ensuremath{\eta}} 

\newcommand{\vfield}{\ensuremath{\boldsymbol{u}}}

\newcommand{\vfieldRed}{\ensuremath{{\boldsymbol{\hat{u}}}}}

\newcommand{\dmdM}{\ensuremath{A}}
\newcommand{\dmdMproj}{\ensuremath{\tilde{\dmdM}}}
\newcommand{\dmdEval}{\ensuremath{\Lambda}}
\newcommand{\dmdeval}{\ensuremath{\lambda}}
\newcommand{\dmdMode}{\ensuremath{\psi}}
\newcommand{\dmdModeProj}{\ensuremath{\tilde{\dmdMode}}}

\newcommand{\dmdRes}{\ensuremath{\overline{\mathcal{R}}}}
\newcommand{\dmdApprox}{\ensuremath{\tilde{\xi}}}

\newcommand{\cutEig}{\ensuremath{c_{\sigma}}}
\newcommand{\impEig}{\ensuremath{c_{\chi}}}

\newcommand{\Ndmd}{\ensuremath{N_d}}

\newcommand{\CFL}{\textit{CFL}} 

\newcommand{\inprod}[2]{\left\langle #1 ,\, #2 \right\rangle}

\renewcommand\Re{\mathrm{Re}\,} 
\renewcommand\Im{\mathrm{Im}\,}

\newcommand{\slphase}{\ensuremath{\phi}}

\newcommand{\vu}{\ensuremath{\mathbf{u}}}

\newcommand{\vuhat}{\ensuremath{\hat{\vu}}}

\newcommand{\vf}{\ensuremath{\mathbf{f}}}

\newcommand{\vx}{\ensuremath{\mathbf{x}}}
\newcommand{\vz}{\ensuremath{\mathbf{z}}}
\newcommand{\vy}{\ensuremath{\mathbf{y}}}
\newcommand{\vxhat}{\ensuremath{\hat{\vx}}}
\newcommand{\vzhat}{\ensuremath{\hat{\vz}}}
\newcommand{\vyhat}{\ensuremath{\hat{\vy}}}

\renewcommand{\d}{\partial}
\newcommand{\oT}{\mathbf{T}}
\newcommand{\oR}{\mathbf{R}}
\newcommand{\oS}{\mathbf{S}}

\newcommand{\sM}{\mathcal{M}}
\newcommand{\sG}{\mathcal{G}}

\title
{
Symmetry-reduced Dynamic Mode Decomposition of Near-wall Turbulence
}
\author{E.\ Marensi\aff{1,2},
    G.\ Yaln\i z\aff{1},
    B.\ Hof\aff{1}
    \and  N.\ B.\ Budanur\aff{1,3}\corresp{\email{nbudanur@pks.mpg.de}}}
\affiliation{\aff{1}Institute of Science and Technology Austria (IST Austria), 
                Am Campus 1, 3400 Klosterneuburg, Austria
            \aff{2}The University of Sheffield, Department of Mechanical Engineering,
            Mappin Street, S1 3JD Sheffield, United Kingdom
            \aff{3}Max Planck Institute for the Physics of Complex Systems (MPIPKS),
            Nöthnitzer Straße 38, 01187 Dresden, Germany}

\pubyear{2022}
\date{\preprintDate}
\begin{document}
\maketitle

\begin{abstract}
Data-driven dimensionality reduction methods such as proper orthogonal
decomposition (POD) and dynamic mode decomposition (DMD) have proven to be
useful for exploring complex phenomena within fluid dynamics and beyond. A
well-known challenge for these techniques is posed by the continuous symmetries,
e.g.\ translations and rotations, of the system under consideration as drifts in
the data dominate the modal expansions without providing an insight into the
dynamics of the problem. In the present study, we address this issue for fluid
flows in rectangular channels by formulating a continuous symmetry reduction
method that eliminates the translations in the streamwise and spanwise
directions simultaneously. We demonstrate our method by computing the
symmetry-reduced dynamic mode decomposition (SRDMD) of sliding windows of data
obtained from the transitional plane-Couette and turbulent plane-Poiseuille flow
simulations. In the former setting, SRDMD captures the dynamics in the vicinity
of the invariant solutions with translation symmetries, i.e.\ travelling waves
and relative periodic orbits, whereas in the latter, our calculations reveal
episodes of turbulent time evolution that can be approximated by a
low-dimensional linear expansion.
\end{abstract}

\section{Introduction}\label{sec/intro}

Turbulence is a strongly nonlinear phenomenon exhibiting chaotic spatio-temporal
behaviour at many scales. Despite its complexity, a certain degree of coherence
is observed and has been studied for many years with the goal of describing
dynamics of turbulent flows in terms of few \textit{coherent structures}
\citep{jimenez2018coherent}. In the context of wall-bounded flows, a
considerable amount of research
\citep[see e.g.][]{hamilton1995regeneration,waleffe1997on,jimenez1999autonomous,schoppa2002coherent}
is devoted to understanding the turbulence-sustaining mechanisms in terms of
\textit{quasi-streamwise vortices}, coherent regions of vortically moving fluid
transverse to the flow direction, and \textit{streaks}, elongated high- or
low-speed modulation of the base flow. Despite the abundant numerical and
experimental evidence supporting the importance of streaks in wall turbulence
and the intuitive physical picture provided by their interactions with vortices,
the definition of a streak is based on experimental observations,
thus inherently subjective \citep{jimenez2018machineaided}. Consequently, one does
not know how much is lost by neglecting the rest of the fluctuations in
turbulent flow.

A complementary, yet mathematically exact, approach to low-dimensionality in
turbulence is provided by the so-termed \citep{waleffe2001exact} {\emph{exact
coherent structures} (ECS), which are unstable time-invariant (self-sustaining)
solutions of the Navier--Stokes equations such as equilibria, travelling waves
and periodic orbits. These correspond to compact low-dimensional objects in the
infinite-dimensional \textit{state space} of all possible flow fields and
influence the dynamics in their vicinity via their stable and unstable manifolds
\citep{gibson2008visualizing,vanveen2011homoclinic,budanur2017relative,
budanur2019geometry,budanur2018complexity,suri2017forecasting,suri2018unstable,suri2019heteroclinic,farano2018computing}.
In other words, together with their stable and unstable manifolds, ECS provide
the intrinsic coordinates that can transiently approximate turbulence. Despite
the importance of ECS being fully established for transitional and
low-Reynolds-number turbulent flows 
\citep[see extensive reviews by][]{kerswell2005recent,eckhardt2007dynamical,
kawahara2012significance,graham2021exact},
tools for computing them become impractical at
higher Reynolds numbers that require many more 
numerical degrees of freedom to resolve in a direct numerical simulation (DNS). 
Thus, the relevance of ECS for complex turbulent flows remains an open 
question. 

The current availability of large data sets, both from experiments and
simulations, and ongoing developments of data-driven modelling tools offer new
avenues for tackling the problem of identifying low-dimensional behaviour
underpinning complex fluid dynamics. Indeed, high-dimensional data can be fed
into data-driven decomposition techniques to gain useful information about the
underlying physical processes \citep{rowley2017model}. Amongst these methods,
dynamic mode decomposition (DMD) \citep{schmid2008dynamic,schmid2010dynamic} has
been successfully applied to many complex fluid systems
\citep[for a comprehensive list see][table 3]{rowley2017model}
with the aim of extracting dynamically important flow features from
time-resolved data. DMD generates a hierarchy of flow fields (DMD modes) and the
associated eigenvalues (DMD eigenvalues) that can be used to approximate the
input data by a linear expansion. Finding a linear modal expansion to describe
strongly nonlinear chaotic fluid dynamics might at first sound like a hopeless
endeavour. However, such an approximation can be found for a finite time,
similar, in spirit, to using a nonlinear invariant solution and its
stable/unstable manifolds to approximate turbulent time evolution in its
neighbourhood. One way of rationalizing this is through the interpretation of
DMD modes as the eigenmodes of the best-fit linear system for the given data set
\citep{kutz2016dynamic}. Another reasoning follows from the correspondence
between DMD and Koopman mode decomposition \citep{rowley2009spectral}, which,
under certain assumptions, states that DMD can be interpreted as a
finite-dimensional approximation to the spectrum of the linear Koopman operator
\citep{koopman1931hamiltonian,mezic2005spectral} that acts on the observables
associated with the dynamical system under consideration. 

In this paper, we present applications of DMD to the data obtained from the DNS
of transitional Couette and turbulent Poiseuille flows in rectangular channels.
The key technical advancement here is the preprocessing of data by symmetry
reduction to eliminate the degeneracies due to streamwise and spanwise
translations, which resolves the well-known shortcomings
\citep{kutz2016dynamic,sesterhenn2019characteristic} of DMD in systems with
continuous symmetries. In such settings, the drifts in the continuous symmetry
directions artificially increase the dimensionality of embeddings that can
reliably capture the dynamics
\citep{rowley2000reconstruction,mendible2020dimensionality,lu2020lagrangian,
sesterhenn2019characteristic,baddoo2021physics-informed}. Furthermore, in
spatiotemporal systems with a continuous flux, such as the Poiseuille flow
considered here, the drifting motion completely dominates the DMD spectra,
obscuring the physically-important dynamics of the system under study. Through
examples in the following, we demonstrate that the \textit{symmetry-reduced
dynamic mode decomposition} (SRDMD) of the channel flows resolves the
aforementioned issues and reveals episodes that can be reliably described by
low-dimensional linear expansions. 

The paper is organized as follows. In \secref{channel}, we introduce channel
flows and our computational setup. We introduce the symmetries of channel flows
and formulate our continuous symmetry reduction method in \secref{symms}. We
summarize the DMD algorithm in \secref{dmd} and then apply it to the
symmetry-reduced DNS data from Couette and Poiseuille flows in \secref{relinv}
and \secref{linear}, respectively. We conclude with a discussion of our results
and the future directions in \secref{conclusions}.

\section{Channel flows and the computational setup}
\label{sec/channel}

We consider flows between two parallel plates in a rectangular
domain $(x, y, z) \in [0, L_x) \times [-1, 1] \times [0, L_z) $, where $x$, 
$y$, $z$ are the streamwise, wall-normal and spanwise directions, respectively.
We take the base-fluctuation decomposition
$\vu_{\text{total}} (\vx, t) = U(y) \vxhat + \vu(\vx, t)$, 
$p_{\text{total}}(\vx, t) = P_x (t) x + p (\vx, t)$
where $U (y) \vxhat$ is the base (laminar) flow, and
$P_x (t)$ is the
spatial mean of the streamwise pressure gradient. Using 
these definitions, the governing Navier--Stokes equations can be written as
 \eq{NS}{
    \d_t \vu =
        - \vu_{\text{total}} \cdot \nabla \vu_{\text{total}} - \nabla p 
        + \Rey^{-1} \nabla^2 \vu 
        + \bracket{\Rey^{-1} \d_y^2 U(y) - P_x} \vxhat\,, 
}
where $\d_a \coloneqq  \d / \d a$ and $\Rey$ is the Reynolds number.
In the following, we consider two base flows, namely $U^{(C)} = y$ (Couette)
and $U^{(P)} = 1 - y^2$ (Poiseuille). In both cases, the fluctuating
velocity fields are periodic in the homogeneous directions, 
i.e.\ $\vu (\vx,t) = \vu (\vx + L_x \vxhat, t) = \vu (\vx + L_z \vzhat, t)$, 
vanish (no-slip) at the walls, i.e.\ $\vu (\vx, t)|_{y=\pm 1} = 0$, and satisfy
the incompressibility condition $\nabla \cdot \vu = 0$.
All of our results to follow
are given for the fluctuations $\vu = [u, v, w](x, y, z)$ from the base flows,
for which we define the $L_2$ inner product 
and the $L_2$ norm respectively as
\eq{norm}{
    \inprod{\vu_1}{\vu_2} 
        \coloneqq \frac{1}{2 L_x L_z}  \int_0^{L_z} \int_{-1}^{1} \int_0^{L_x} 
        \vu_1 \cdot \vu_2 \, \mathrm{d}x\, \mathrm{d}y\, \mathrm{d}z \, 
    \quad\mbox{and}\quad \|\vu\| = \sqrt{\inprod{\vu}{\vu}}\,.
}

We utilize Channelflow 2.0 \citep{Channelflow2} for the numerical integration of
\eqref{NS} in computational domains, properties of which are summarized
in \tabref{setup}. In all simulations, we use dynamically-adjusted time steps 
so that the Courant--Friedrichs--Lewy number (\CFL) satisfies $0.15 \le
\CFL < 0.3$. 
Our first Couette domain W03 is identical to that of \cite{waleffe2003homotopy}
and the second domain HKW is  based on 
\cite{hamilton1995regeneration}, see \tabref{setup}.
For the latter, we adopt the resolution used in
\cite{viswanath2007recurrent}, which is higher than the one used in 
\cite{hamilton1995regeneration}.
For Poiseuille flow, we chose $\Rey = 2000$
    and $\Rey = 5000$, which we refer to as P2K and P5K, respectively, 
    in \tabref{setup} and hereafter.
We determined the spatial resolutions such that the energy stored in
the Fourier/Chebyshev modes with the highest wave numbers are 
at least six orders of magnitude smaller than those in the lowest ones at all times. 
All of our domains are ``minimal'' in the sense that if the spanwise extent is reduced, 
the simulations quickly laminarize. The Couette simulations are carried under
the constraint $P_x = 0$, 
hence the fluid flux 
$Q_x (t) = \int_{-1}^{1} \int_0^{L_z} \vu \cdot \vxhat\, \mathrm{d}y\, \mathrm{d}z\,$ varies
instantaneously.
In contrast, we simulate Poiseuille flow under the
constraint $Q_x = 4 L_z / 3$,
which leaves $P_x (t)$
fluctuating. Additionally in the Poiseuille case, we impose symmetry invariance
with respect to the midplane on the velocity fields, which restricts the
dynamics into a lower-dimensional flow-invariant subspace without altering wall
friction and the Reynolds stresses near the wall. 
For the Poiseuille systems P2K and P5K, we estimate the friction Reynolds 
numbers and the channel dimensions in wall units \citep{pope2000turbulent} as 
$(\Rey_{\ta}, L_x^+, L_z^+) \approx (98, 280, 123)$ and 
$(\Rey_{\ta}, L_x^+, L_z^+) \approx (205, 280, 120)$ respectively.
Note that our spanwise domain length is slightly larger than
the minimal flow unit $L_z^+\approx100$ established in
\citep{jimenez1991minimal}. We suspect that this is due to our symmetry
constraint which does not allow for single-wall 
localization of turbulent structures
that was observed by Jimen\'ez and Moin at this $\Rey$. 

\tab{setup}{The laminar flows $U(y)$, the domain lengths $L_x$ and $L_z$, 
the grid dimensions $N_x$, $N_y$, $N_z$, and the additional constraints 
of the computational setups. 
Here W03 and HKW correspond to the Couette domains of 
 \cite{waleffe2003homotopy} and \cite{hamilton1995regeneration},
 respectively, while P2K and P5K correspond to the
  Poiseuille systems at $Re=2000$ and 5000, respectively.
}{
    c c c c c
}{
            & W03            & HKW            & P2K                                  &P5K 
}{          
$U(y)$      & $y$            & $y$            & $1 - y^2$                          & $1 - y^2$ \\
$\Rey$      & $400$          & $400$          & $2000$                             & $5000$ \\
$L_x$       & $2 \pi / 1.14$ & $2 \pi / 1.14$ & $2 \pi / 2.2$                      & $2 \pi / 4.6$ \\
$L_z$       & $2 \pi / 2.5$  & $2 \pi / 1.67$ & $2 \pi / 5$                        & $2 \pi / 10.7$ \\
$N_x$       & $48$           & $48$           & $64$                               & $64$ \\
$N_z$       & $35$           & $48$           & $48$                               & $48$ \\
$N_y$       & $48$           & $65$           & $97$                               & $193$ \\
Constraints & $P_{x}$=0    & $P_{x}$=0        & $\dot{Q}_{x}$=0, $\oR_y \vu$=$\vu$ & $\dot{Q}_{x}$=0, $\oR_y \vu$=$\vu$ 
}

\section{Symmetries and symmetry reduction}
\label{sec/symms}

Both Couette and Poiseuille systems are equivariant under the translations
\eq{trans}{
\oT (\de x, \de z) [u, v, w](x, y, z) = [u, v, w](x - \de x, y ,z - \de z)  \,,
}
where $\de{x} \in [0, L_x)$ and $\de{z} \in [0, L_{z})$, and the reflection  
\eq{refz}{ 
    \oR_z [u, v, w](x, y, z) =
        [u, v, - w](x, y, - z)\,.        
}
Additionally, Poiseuille flow admits the equivariance under the
reflection 
\eq{refy}{
    \oR_y [u, v, w](x, y, z) = 
    [u, - v, w](x, - y, z)
}
with respect to the midplane; and plane-Couette flow is equivariant 
under the simultaneous reversal of streamwise and wall-normal directions,
i.e.\
\eq{refxy}{
    \oR_{xy} [u, v, w](x, y, z) = [-u, -v, w](-x, -y, z)\,. 
}

For the present study, the equivariance of a flow under a
symmetry group $\sG$
has two important consequences
\citep{golubitsky1985singularities,chossat2000methods}: (i) If $\vu (\vx, t)$
for $t \in [t_i, t_f]$ is a trajectory of the system, then so is 
$\oS \vu (\vx, t)$ where $\oS \in \sG$;
(ii) 
If $\vu (\vx, 0)$ is invariant under $\oS \in \sG$ satisfying 
$\oS \vu (\vx, 0) = \vu (\vx, 0)$, then its
forward-time evolution remains invariant under $\oS$, i.e.\ $\oS \vu (\vx, t) = \vu
(\vx, t)$ for $t > 0$. We make use of (ii) when restricting our study of
Poiseuille flow into the space of solutions that are invariant under
\eqref{refy}. Due to the presence of continuous symmetries, (i) effectively 
implies that the generic solutions of Couette and Poiseuille flows have 
infinitely-many symmetry copies due to translations and their combinations
with various reflections. 

\cite{sirovich1987turbulenceb} showed that if a data set of flow states
is symmetric under a continuous translation, then its proper orthogonal
decomposition (POD) results in modes that align with Fourier modes in the
homogeneous directions, carrying no information about the physics of the
system. As a remedy, \cite{rowley2000reconstruction}
suggested reducing the symmetry degree of freedom prior to the POD of the data
obtained from a system with translation symmetry. Their method relied on an
experimentally-chosen template to which the simulation data is matched. As noted
by the authors themselves, such a symmetry reduction method has a finite domain
of applicability, the boundary of which is set by the singularity of the
so-called reconstruction equation. 
Recently, the difficulties posed by continuous symmetries for dimensionality 
reduction have also received attention in DMD and machine learning 
literature and several new techniques to address them were proposed.
\cite{sesterhenn2019characteristic} suggested a space-time rotation
that
can be 
employed at a characteristic group velocity to improve the performance of 
DMD in drift-dominated systems. 
\cite{lu2020lagrangian} introduced 
the Lagrangian DMD which requires one to co-evolve the solution grid along 
with the scalar fields. 
In the physics-informed DMD developed by 
\cite{baddoo2021physics-informed}, the DMD matrix that provides 
the best-fit 
linear system to the data is constrained to the space of matrices that 
commute with the symmetry operators. 
Finally, \cite{kneer2021symmetryaware} utilize the 
so-called spatial translation 
networks to perform template matching akin to that of 
\cite{rowley2000reconstruction}. 
Each of these methods come with a new set of technical difficulties and it is 
unclear whether they are practical for the three-dimensional complex fluid 
flows that we consider here.
In the following, we avoid these difficulties by taking an approach 
similar to that of \cite{rowley2000reconstruction}
and formulate a symmetry reduction method for preprocessing channel flow data
prior to its DMD. Differently from \cite{rowley2000reconstruction}, our method
yields a symmetry reduction for all dynamics of interest. 

\cite{budanur2015reduction} showed that a polar coordinate
transformation in the Fourier space of a spatially extended system can be
interpreted as a \textit{slice}, that is, a codimension-1 manifold in the state
space where each set of translation-equivalent states is
represented by its unique intersection with this manifold. On applications to
the Kuramoto--Sivashinsky system, \cite{budanur2015reduction}
demonstrated that such a \textit{first Fourier mode slice} can be used to reduce
the translation symmetry of the flow for all dynamics of interest. Later, the
method was successfully adapted to two-dimensional
Kolmogorov \citep{farazmand2016adjointbased,hiruta2017intermittent} and
three-dimensional pipe
\citep{willis2016symmetry,budanur2017relative,budanur2018complexity} flows; see
\cite{budanur2015periodic} for a pedagogical introduction. Here, we formulate this 
method to flows in rectangular channels. We begin by defining the slice templates
\eqarr{
    \vuhat'_x &\coloneqq& \vf_x (y) \cos (2 \pi x / L_x)\,, \label{slicetempx} \\
    \vuhat'_z &\coloneqq& \vf_z (y) \cos (2 \pi z / L_z)\,, \label{slicetempz}
}
where $\vf_x (y)$ and $\vf_z (y)$ are to-be-specified vector-valued 
functions of the wall-normal coordinate only.  
Let $\vu$ be a solution and the set 
$\sM_{\vu}^x = \{ \oT (\de{x}, 0) \vu \,|\, \delta x \in [0, L_x) \}$ 
be formed by $\vu$ and its streamwise-translation copies. The key idea behind 
the first Fourier mode slice is the observation that any nonzero projection
of $\sM_{\vu}^x$ onto the plane spanned by
$\vuhat'_x$ and its quarter-domain shift $\oT(L_x/4, 0)
\vuhat'_x = \vf_x (y) \sin (2 \pi x / L_x)$ is of circular shape. Thus, a 
transformation that fixes the polar angle 
\(
\slphase_x \coloneqq \arg
\left(\inprod{\vu}{\vuhat'_x} + i \inprod{\vu}{\oT (L_x/4, 0) \vuhat'_x }\right)\,
\)
can be used to reduce the translation symmetry. 
Following analogous observations, we define
\(
\slphase_z \coloneqq \arg
\left(\inprod{\vu}{\vuhat'_z} + i \inprod{\vu}{\oT (0, L_z/4) \vuhat'_z }\right)\,
\)
and the symmetry-reducing transformations
\eqarr{
    S_x (\vu) &\coloneqq&
        \oT \pa{- \frac{\slphase_x L_x}{2 \pi}, 0} \vu \,, \label{symredx} \\
    S_z (\vu) &\coloneqq&
        \oT \pa{0, - \frac{\slphase_z L_z}{2 \pi}} \vu \label{symredz} \,.  
}
Noting that the slice templates $\vuhat'_x (x, y)$ \eqref{slicetempx} and 
$\vuhat'_z (y, z)$ \eqref{slicetempz} do not depend on the $z$ and $x$ coordinates
respectively and the translations in $x$ and $z$ directions commute,
we reduce the streamwise and spanwise translations simultaneously by
simply applying \eqref{symredx} and \eqref{symredz} consecutively as 
\eq{symred}{
    \vuhat = S (\vu) = S_z (S_x (\vu)) \, .  
}

Until now, we left the wall-normal dependence of the template functions
(\ref{slicetempx}, \ref{slicetempz}) unspecified. In order to clarify this final
point, let us first give a geometric interpretation of
continuous symmetry reduction. Since symmetry reduction eliminates two continuous 
translation degrees of freedom, the symmetry-reduced velocity fields $\vfieldRed (t)$ 
are confined to a submanifold in the state space with two dimensions less than that
accommodating the original velocity fields $\vfield (t)$. This information, however, 
is not lost and can be recovered as long as one keeps track of the slice phases 
$\slphase_x (t)$  and $\slphase_z(t)$. \cite{rowley2000reconstruction} showed 
that these phases can also be obtained by integrating the reconstruction equations 
\eqarr{ 
    \dot{\slphase}_{x} (t) = \pa{\frac{2 \pi}{L_x}} 
        \frac{\inprod{\d_{x} \vuhat'_{x}}{\d_{t} \vu|_{\vu =\vuhat (t)}}}{ 
              \inprod{\d_{x} \vuhat'_{x}}{\d_{x} \vuhat (t)}}\,,  
    \label{slicephasex} \\ 
    \dot{\slphase}_{z} (t) = \pa{\frac{2 \pi}{L_z}}
        \frac{\inprod{\d_{z} \vuhat'_{z}}{\d_{t} \vu|_{\vu = \vuhat (t)}}}{ 
              \inprod{\d_{z} \vuhat'_{z}}{\d_{z} \vuhat (t)}}\,.
    \label{slicephasez}        
}
Note that these phase velocities diverge if the
denominators of the reconstruction equations vanish, at which point
our symmetry reduction method would suffer a discontinuity. It is 
straightforward to confirm that these denominators are proportional 
to the amplitudes of the projections of the flow state $\vu$ onto 
the plane spanned by the respective slice template and their 
half-domain shift. In other words, 
\eqarr{
    \inprod{\partial_{x} \vuhat'_{x}}{ 
        \partial_{x} \vuhat (t)} &\propto& 
        \sqrt{\inprod{\vu}{\vuhat'_x}^2 + \inprod{\vu}{\oT (L_x/4, 0) \vuhat'_x }^2} 
        \,, \label{vuxhatproj} \\
    \inprod{\partial_{z} \vuhat'_{z}}{ 
        \partial_{z} \vuhat (t)} &\propto& 
        \sqrt{\inprod{\vu}{\vuhat'_z}^2 + \inprod{\vu}{\oT (0, L_z/4) \vuhat'_z }^2} 
        \,, \label{vuzhatproj}
}
thus, as long as these projections onto the template-planes do not 
vanish, the right-hand sides of the
reconstruction equations (\ref{slicephasex}, \ref{slicephasez}) 
remain finite. With this in mind, we determine $\vf_x (y)$ and $\vf_z (y)$ as 
follows to maximize the projection amplitudes
\eqref{vuxhatproj} and \eqref{vuzhatproj} 
for turbulent trajectories. Let 
\eqarr{
    \vf_x (y) &=& \sum_{n = 0}^{n_f} T_n(y)
        [c_x^{(x,n)} \vxhat + c_x^{(y,n)} \vyhat + c_x^{(z,n)} \vzhat]\,,
        \label{vfx} \\
    \vf_z (y) &=& \sum_{n = 0}^{n_f} T_n(y)
        [c_z^{(x,n)} \vxhat + c_z^{(y,n)} \vyhat + c_z^{(z,n)} \vzhat]\,, 
        \label{vfz} 
}
where $T_n (y)$ are the Chebyshev polynomials of the first kind and  
$c_{j}^{i,n}$ ($i \in \{x, y, z\},\, j \in \{x, z\},\, n \in \{0, 1, \ldots, n_f\}$) 
are the coefficients which we determine by maximizing 
\eqarr{
    \mathcal{J}_x &=& \sum_{k = 0}^{K}  
        \inprod{\vu(k \de t)}{\vuhat'_x}^2 
        + \inprod{\vu(k \de t)}{\oT (L_x/4, 0) \vuhat'_x }^2\,, 
        \label{slicecostx} \\
    \mathcal{J}_z &=& \sum_{k = 0}^{K}  
        \inprod{\vu(k \de t)}{\vuhat'_z}^2 
        + \inprod{\vu(k \de t)}{\oT (0, L_z/4) \vuhat'_z }^2\,, 
        \label{slicecostz} 
}
for a turbulent trajectory $\{\vu (t), t \in [0, K \delta t]\}$,
under the constraints
$\|\vuhat'_x\| = 1$ and $\|\vuhat'_z\| = 1$.
The necessity of these unit-norm constraints can be understood 
by observing that \eqref{slicephasex} and \eqref{slicephasez} 
are invariant under the scaling of $\vuhat'_x$ and $\vuhat'_z$ 
by a constant. Thus, if we do not apply these constraints, then
the prescribed optimization would diverge by arbitrarily increasing 
the template amplitudes, without a reduction in phase fluctuations.
In the case of Poiseuille flow we have the additional constraint
that the slice templates be $\oR_y$-symmetric like the underlying flow.
The cost functions 
\eqref{slicecostx} and \eqref{slicecostz} are the sums of squares of  
projection amplitudes (right-hand sides of \eqref{vuxhatproj} and 
\eqref{vuzhatproj}).
For each domain that we study, we 
determine \eqref{vfx} and \eqref{vfz} truncated at $n_f = 7$ using a  
single turbulent trajectory sampled at steps of $\ts = 0.1$. 
The resulting $\vf_x (y)$ and $\vf_z(y)$ are plotted in \figref{templates} and the
slice templates can be 
downloaded from
\cite{yalniz2022srdmd}.
The phase velocities (\ref{slicephasex}, \ref{slicephasez}) are plotted in 
\figref{slicephases}. As shown, although the phase speeds exhibit occasional
fast episodes, they remain finite throughout our simulations.   
Note that in \figref{slicephases}, all phase velocities but $\dot{\slphase}_x$ 
of plane-Poiseuille flow fluctuate about 
zero. This is a consequence of
Poiseuille flow's broken reflection symmetry in the streamwise direction 
due to the presence of a nonzero mean pressure gradient. 
Intuitively, one can understand this by considering the presence of the net
drift due to the nonzero bulk velocity $U_b$ in Poiseuille flow. 
However, it is also important to note that $\dot{\slphase}_x$ in Poiseuille flow
is \emph{not} equal to $2 \pi U_b / L_x$, neither is any other phase velocity 
equal to $0$, the bulk velocity at their respective directions, at all times:
Phase velocities vary instantaneously and match
the corresponding bulk velocities only when averaged over long periods.

\fig{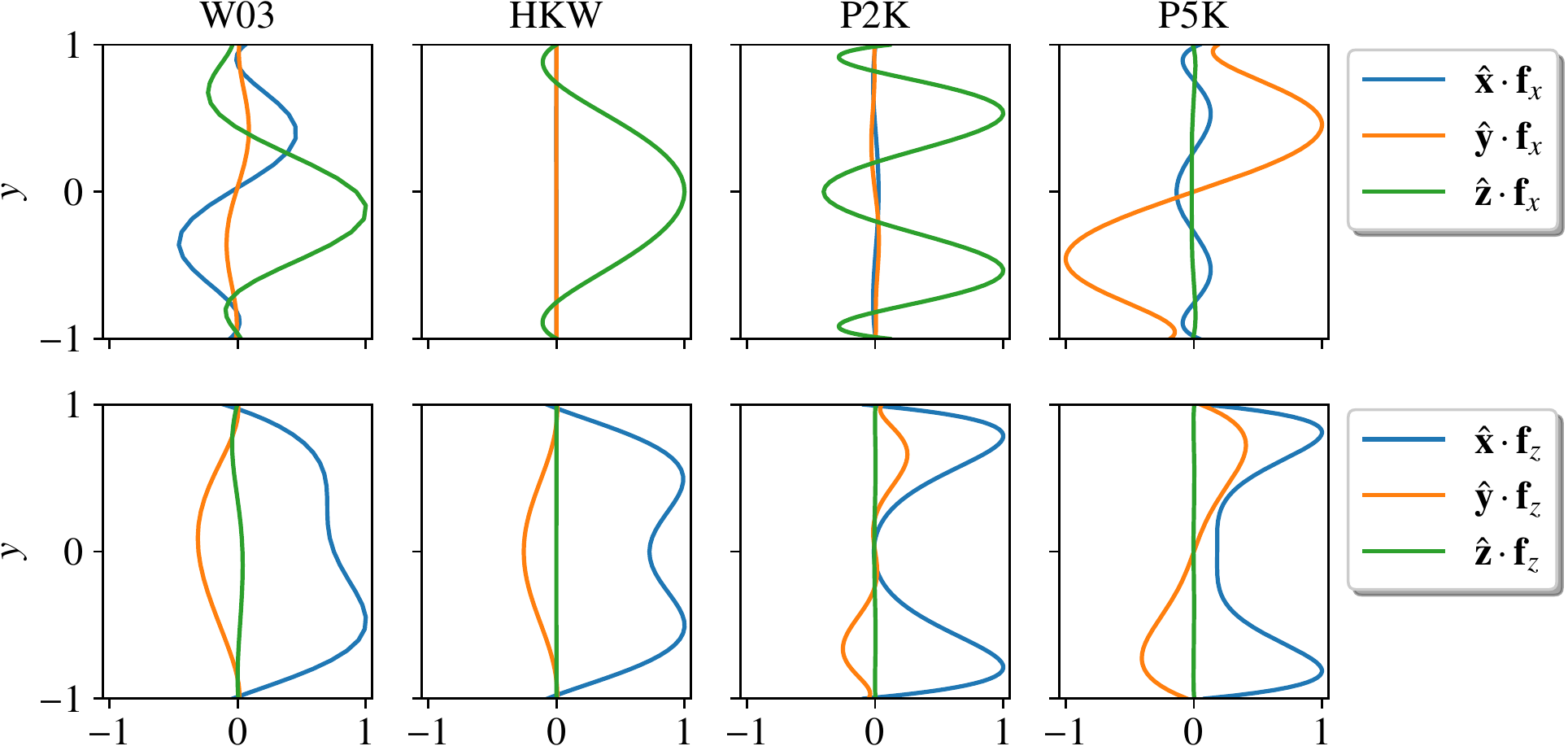}{
    Wall-normal dependencies of the slice templates.
    Columns correspond to the domains studied,
    with the domain name noted on the top.
    Each $\vf_x$ and $\vf_z$ was normalized with
    $\max |\vf_x|$ and $\max |\vf_z|$, which does not affect slicing,
    in order for the plots to share the horizontal axes.
}

\fig{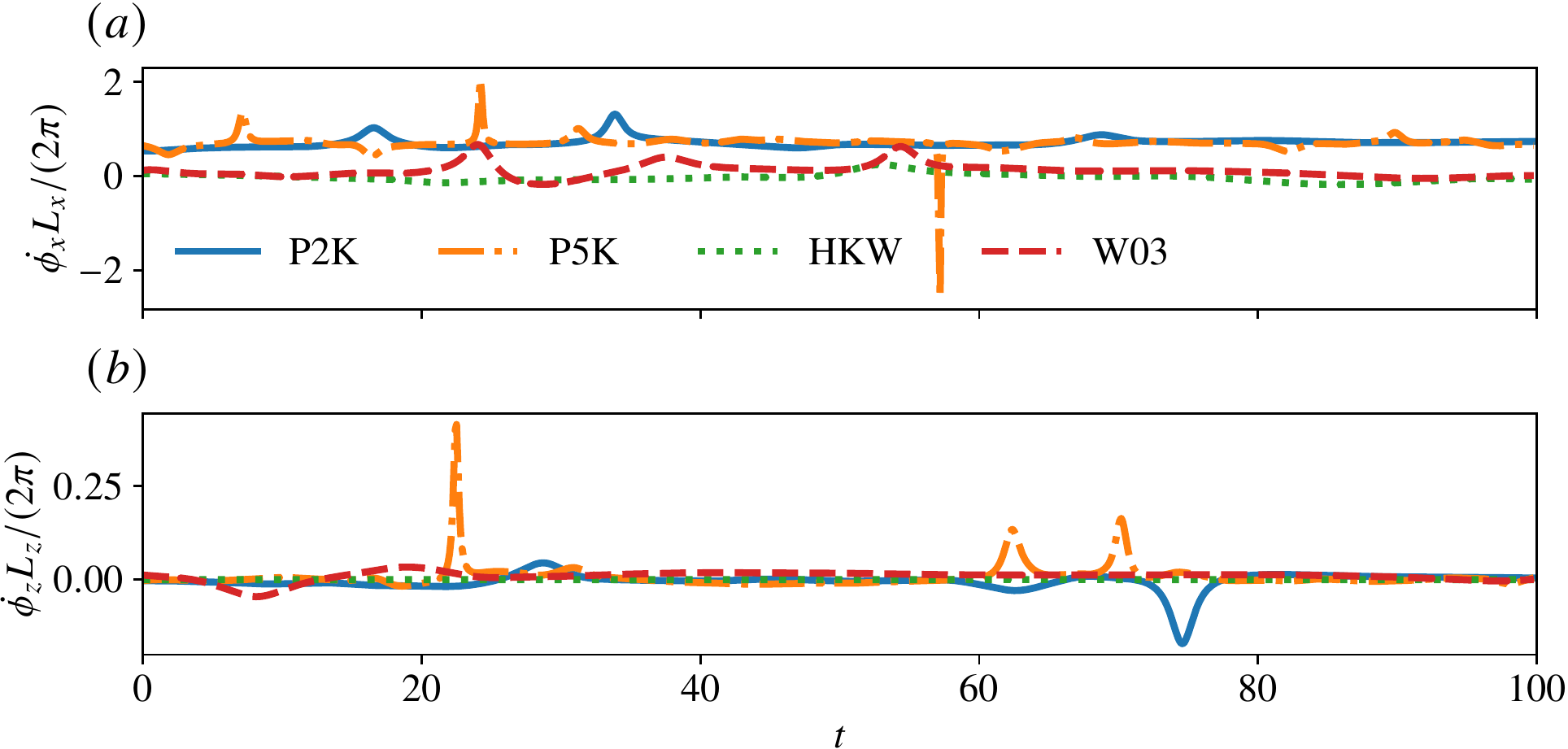}{
    Finite-difference approximations 
    \(\dot{\slphase}_{x, z} (t) 
    \approx 
    (\slphase_{x, z} (t + \sldt) - \slphase_{x, z} (t))/\sldt\)
    with \(\sldt = 0.1\) to the slice phase velocities 
    $\dot{\slphase}_x$ (a) and 
    $\dot{\slphase}_z$ (b)
    corresponding to turbulent trajectories in simulation domains considered.
    $\dot{\slphase}_{x, z}$ are normalized by
    $2\pi/L_{x,z}$ to present the different domains together.
}

As we shall further explain in \secref{relinv} through an example, the
episodes with fast phase velocities in \figref{slicephases} correspond to
those at which chaotic trajectories have relatively small projection
amplitudes (\ref{vuxhatproj}, \ref{vuzhatproj}). Observing this, one might
suggest the temporal minimum of these projections as a cost function to maximize
as opposed to the sums of squares (\ref{slicecostx},\, \ref{slicecostz}).
Although we experimented with such a cost function,  
ultimately we opted
against it because the optimization problem of maximizing a temporal minimum
is non-differentiable, thus significantly more complex, since the instance
of the minimum jumps during the optimization 
procedure. In addition to its
computational simplicity, another motivation to use the cost functions
\eqref{slicecostx} and \eqref{slicecostz} is that we also want the episodes
with fast phase oscillations to be infrequent.
Note that in all the domains we
considered, the wall-normal dependence of the spanwise slice 
template $\vuhat_z$ 
has the largest contribution from the streamwise fluctuations
as shown in \figref{templates} (bottom row). 
Remembering that all of our computational domains are minimal flow units 
\citep{jimenez1991minimal}, we
interpret our optimal slice templates as those that fix the
spanwise positions of streaks, since the minimal flow units are
characterized by the presence of a single pair of fast/slow streaks, 
which appear predominantly in the first spanwise Fourier mode 
of streamwise velocity. Conversely, the
streamwise contribution to the streamwise slice templates  
is much smaller (\figref{templates}, top row) since in this 
direction, streaks make the
largest contribution to the zeroth streamwise Fourier mode
of streamwise velocity. 
Physically, we expect the symmetry reduction procedure to eliminate the 
drifts of the flow structures whose first Fourier mode components  
align with the slice templates. It should be noted and can also be seen 
in Supplementary Movies 1 \& 2
that drifts with 
respect to these structures are 
still present in the symmetry reduced time evolution 
because in a turbulent 
flow fluctuations are advected at the local mean velocity which varies 
within the domain. In other words, symmetry reduction does not eliminate 
all drifts in the velocity fluctuations, but rather eliminates the translation 
degrees of freedom in the data by finding a representative state for each set 
of states that can be mapped to one another via 
symmetry operations.

\section{Symmetry-reduced dynamic mode decomposition}
\label{sec/dmd}
Let $\xi (t)$ be the $n$-dimensional symmetry-reduced state vector 
corresponding to the fluid state at time $t$, $\flow{t}{\xi}$ be the 
finite-time flow induced by the DNS and symmetry reduction, 
and $\inprod{\xi_1}{\xi_2}$ and $\| \xi \|$ denote the $L_2$ inner 
product and norm, respectively, of the corresponding velocity fields 
as defined in \eqref{norm}.
Let $\xi_k$ and $\xi'_k$ be a pair of 
snapshots that are separated by time $\delta t$, i.e.\
$\xi'_{k} = \flow{\delta t}{\xi_{k}}$. 
Defining the $n \times m$ ($n \gg m$)
data matrices 
$\Xi  \coloneqq \left[ \xi_0, \xi_1, \ldots, \xi_{m-1} \right]$ and
$\Xi' \coloneqq \left[ \xi'_0, \xi'_1, \ldots, \xi'_{m-1} \right]$,  
we consider the linear approximation $ \Xi' \approx \dmdM \,\Xi$, where $\dmdM$ 
is an $n \times n$ matrix. The best fit (in $L_2$ sense) to this approximation 
is given by $\dmdM = \Xi' \Xi^\dagger$, where $\dagger$ denotes the 
Moore--Penrose pseudoinverse. We adopt the standard DMD algorithm 
\citep{tu2014on,kutz2016dynamic}, which approximates the eigenvalues and 
eigenvectors of $\dmdM$ without explicitly computing it as follows. Let 
$\Xi \approx U \Sigma V^*$ denote the rank-$r$ ($r < m$) singular value 
decomposition (SVD) approximation of $\Xi$, where 
\(
    U \in \mathbb{C}^{n \times r},\, 
    \Si \in \mathbb{C}^{r \times r},\,
    V \in \mathbb{C}^{m \times r} 
\)
and $^*$ indicates the Hermitian transpose.  
Noting that the columns of $U$ are the POD modes, we 
can rewrite the best-fit linear operator 
and its $r \times r$ projection onto the POD space as 
$\dmdM = \Xi' V \Sigma^{-1} U^*$ 
and $\dmdMproj = U^* \dmdM U = U^* \Xi' V \Sigma^{-1} $, 
respectively. Finally, we compute the eigenvalues $\dmdEval_j$ and eigenvectors 
$\dmdModeProj_j$ of $\dmdMproj$, from which we obtain the SRDMD modes as 
$\dmdMode_j = \Xi' V \Sigma^{-1} \dmdModeProj_j$. Hereafter, we refer to 
$\dmdEval_j$ as the ``SRDMD multipliers'' and 
$\dmdeval_j \coloneqq \ln ( \dmdEval_j) / \delta t$ as the ``SRDMD exponents''.
 With these definitions, we can now write the SRDMD approximation
 of the time-evolution as
\eq{dmdApprox}{
    \dmdApprox(t) = \sum_{j = 0}^{\Ndmd - 1} c_j \dmdMode_j
		e^{\dmdeval_j t} \approx \xi(t) \,, 
}
where $c_j$ are the SRDMD coefficients and $\Ndmd \le r$ is the
number of SRDMD modes used to reconstruct the velocity field.
 Following \cite{page2019koopman}, we set
the coefficients $c_j$ 
as those that minimize the cost
function
\eq{cost}{
\mathcal{J}(c_0, c_1, \ldots, c_{\Ndmd - 1}) =  \sum_{k = 0}^{m -
    1} \| \xi(t_k)
-\dmdApprox(t_k) \|^2 \,. 
} 
In the following, we refer to the SRDMD mode $\ps_0$ with 
$\La_j \approx 1$ ($\la_j \approx 0$) as the ``marginal mode''
and sort the rest according to their 
normalized spectra \citep{tu2014on} in descending 
$|\La_j |^{m} \| c_j \psi_j \|$. 
Note that ordering the SRDMD 
modes in this way amplifies (penalizes) those that grow (decay)
by multiplying them with their respective multiplier raised to the power
$m$.

We compute the SVD of $\Xi$
using the method of snapshots \citep{sirovich1987turbulence} 
and follow \cite{holmes1996turbulence,sirovich1989chaotic} 
to truncate it such that a
sufficiently large fraction \cutEig{} of the total energy is captured
and no neglected mode contains, on average, more than a small
fraction \impEig{} of the energy contained in the first mode.
Namely
\begin{equation}
\sum_{i=0}^{r-1} \sigma_i^2 > \cutEig\, \sum_{i=0}^{m-1} \sigma_i^2  \quad\mbox{and}\quad
\frac{1}{m - r}\sum_{i=r}^{m - 1}\sigma_i^2 < \impEig\, \sigma_0^2\,,
\end{equation}
where $\sigma_i$ are the singular values.
For all of our results to follow, we set 
$\cutEig = 0.9999$ and $\impEig=0.001$ which we
determined by ensuring that 
higher-rank truncations
do not alter the leading SRDMD exponents
in the first two digits. 

Note that the above-summarised formulation of DMD does not 
necessitate
the data points $\xi_0, \xi_1, \xi_2, \ldots$ to be uniformly distributed in
time. The only requirement is for the snapshot separation time $\de t$ to
be fixed across all snapshot pairs $(\xi_k, \xi'_k)$. Therefore, one can 
increase the number of data points corresponding to a time interval by sampling 
it at a time step $\ts = \de t / n$, where $n \in \mathbb{Z}^+$. In our plane-Couette
examples of \secref{relinv}, 
we make use of this property by choosing $\ts = \de t /  10$, whereas in our 
Poiseuille flow demonstrations of \secref{channel}, we use uniformly distributed 
samples separated by $\de t$.

\section{Relative Invariant Solutions and their SRDMD}
\label{sec/relinv}

\cite{page2019koopman,page2020searching} demonstrated how DMD 
captures dynamics of the plane-Couette flow near 
simple equilibria and 
periodic orbits defined by 
\eq{invariants}{
    \vu_\mathrm{eq}(t) = \vu_\mathrm{eq}(0) \quad \mbox{and} \quad
    \vu_\mathrm{po}(t + T_\mathrm{po}) = \vu_\mathrm{po}(t) \,, 
} respectively. The equilibria of the 
plane-Couette flow belong to the flow-invariant subspaces of $\oS_1 \oR_z$ and 
$\oS_2 \oR_{xy}$, where 
$\oS_{1}$ and $\oS_{2}$ are some elements of plane-Couette flow's 
symmetry group. 
Invariance of these solutions under the symmetries involving reflections 
$\oR_z$ and $\oR_{xy}$ restricts
their dynamics to the space of ``non-drifting'' velocity fields, 
see \cite{gibson2009equilibrium} for details.
    Besides the equilibria, these flow-invariant subspaces can also 
    accommodate periodic orbits. 
    Alternatively, the periodic orbits can be formed by two periods 
    of a ``preperiodic'' \citep{budanur2016unstable} solution defined by
    $\vu_\mathrm{ppo}(t + T) = \oS_r \vu_\mathrm{ppo} (t)$, where $\oS_r \in \sG$ 
    satisfies $\oS_r^2 = \identity$.
When such symmetries are not present, the 
generic solutions of plane-Couette flow exhibit streamwise and spanwise 
drifts. The simplest invariant solutions with drifts are the relative 
equilibria that satisfy
\eq{releq}{
    \vu_\mathrm{req}(t) = \oT(c_x t, c_z t) \vu_\mathrm{req}(0) \,, 
}
where $c_x$ and $c_z$ are phase velocities; and the drifting counterpart 
of the periodic orbits are the relative periodic orbits defined by
\eq{rpo}{
    \vu_\mathrm{rpo}(t + T_\mathrm{rpo}) 
    = \oT(\De x_\mathrm{rpo}, \De z_\mathrm{rpo}) \vu_\mathrm{rpo} (t) \,. 
}

As our first demonstration of SRDMD, we apply it to trajectories in the vicinity
of the relative equilibrium $\TW_3$, a travelling wave originally found by 
\cite{gibson2009equilibrium} in the W03 domain.\footnote{ 
    The data for this solution is available in the \texttt{channelflow.org} 
    database.}
\Figref{TW3}(a) shows the SRDMD exponents (blue crosses) computed using five
trajectories initiated as random perturbations to $\xi_{\TW_3}$ with
perturbation amplitudes equal to $10^{-2} \|\xi_{\TW_3}\|$.   
We integrated each of these trajectories in time
and sampled the states at time steps of $\ts=0.1$
in the time interval $t=[10,80]$ in which the dynamics 
was found to be approximately linear. 
To construct the data matrices $\Xi$ and $\Xi'$, we chose a 
separation time $\delta t=1$ between the corresponding snapshots $(\xi, \xi')$ and
we randomly selected 200 pairs of snapshots out of the 
 $(T_w-\delta t)/\ts +1$ possible samples
 from each of the five trajectories,
 where the window length is $T_w=70$.
Using the resulting SRDMD modes, we computed the 
best-fit coefficients
$c_j^{(k)}$
as explained in \secref{dmd} for each trajectory $k = 1,2, \ldots, 5$
and ordered 
the exponents 
in descending
$(1/5) \sum_{k=1}^{5}|\La_j |^{m} \| c_j^{(k)}  \psi_j \|$, 
i.e. according to the trajectory-averaged normalized spectrum.
For comparison, \figref{TW3}(a) also shows the linear 
stability spectrum of $\TW_3$, which we approximated via Arnoldi iteration 
\citep{trefethen1997numerical} using its \texttt{channelflow} implementation. 
As shown, the SRDMD exponents yield an approximation to the 
leading (ordered in descending real parts) linear stability 
eigenvalues of the travelling wave.
Specifically, the four unstable eigenvalues with 
$\Re \la > 0$ of $\TW_3$ are captured very well by SRDMD, 
whereas the stable part of the spectrum can only be partially 
observed among the SRDMD eigenvalues and 
a spurious complex-conjugate SRDMD stable mode with $\Re \la < -0.05$ 
is also present in \figref{TW3}(a).
In contrast, when we repeat this computation 
without symmetry reduction, we find that 
all non-marginal DMD modes 
simply lie at the drift frequency and its multiples, 
as
seen in \figref{TW3}(d), and thus carry
no information about the dynamics in the vicinity of the travelling wave.

To further illustrate how SRDMD captures dynamics in the vicinity of
a travelling wave, we perturbed $\TW_3$ in the directions of the
eigenvectors corresponding to the eigenvalues
$\la_3$ and $\la_9$
(counting eigenvalues starting from the most unstable $\la_1$), 
and computed
SRDMD approximations 
to these trajectories using four modes, i.e. $N_d = 4$. 
Each of these calculations effectively resulted 
in three-dimensional SRDMD approximations with the 
fourth mode amplitude being negligible. Indeed, after symmetry reduction,
one neutral mode along with a pair of complex conjugate is sufficient
for capturing spiral in/out dynamics. In \figref{TW3}(b,c) these
trajectories and their SRDMD approximations are visualized as projections onto
the hyperplanes spanned by the real and imaginary parts of the leading non-marginal 
SRDMD modes, centred around their respective marginal modes.
In these and all of our state space projections to follow, the  
axes correspond to 
$p_i = \inprod{\xi (t) - \ps_0}{\ps_i}$ and
$p_i = \inprod{\tilde{\xi} (t) - \ps_0}{\ps_i}$ for DNS trajectories and 
their SRDMD approximations, respectively.
As shown, SRDMD nicely 
captures the spiral-out/in dynamics of 
these unstable/stable neighbourhoods.

In order to demonstrate how the drifting motion of the travelling wave obscures
the dynamics when the continuous symmetries are not taken into consideration, we
repeated our spectrum calculation and approximations to the unstable/stable
subspaces without reducing the continuous symmetries.
As shown in \figref{TW3}
(d), when the drifts are not eliminated, the DMD exponents show no resemblance
to the spectrum of $\TW_3$ and the individual DMD approximations are completely
dominated by the drifts as indicated by the approximately-circular projections
in \figref{TW3}(e,f). Although DMD still approximates the 
trajectories shown in \figref{TW3}(e,f), essentially the only information carried 
in these projections is that of the drifts and only after symmetry reduction 
in \figref{TW3}(b,c) can one see that these 
trajectories belong to different dynamical regimes.

\fig{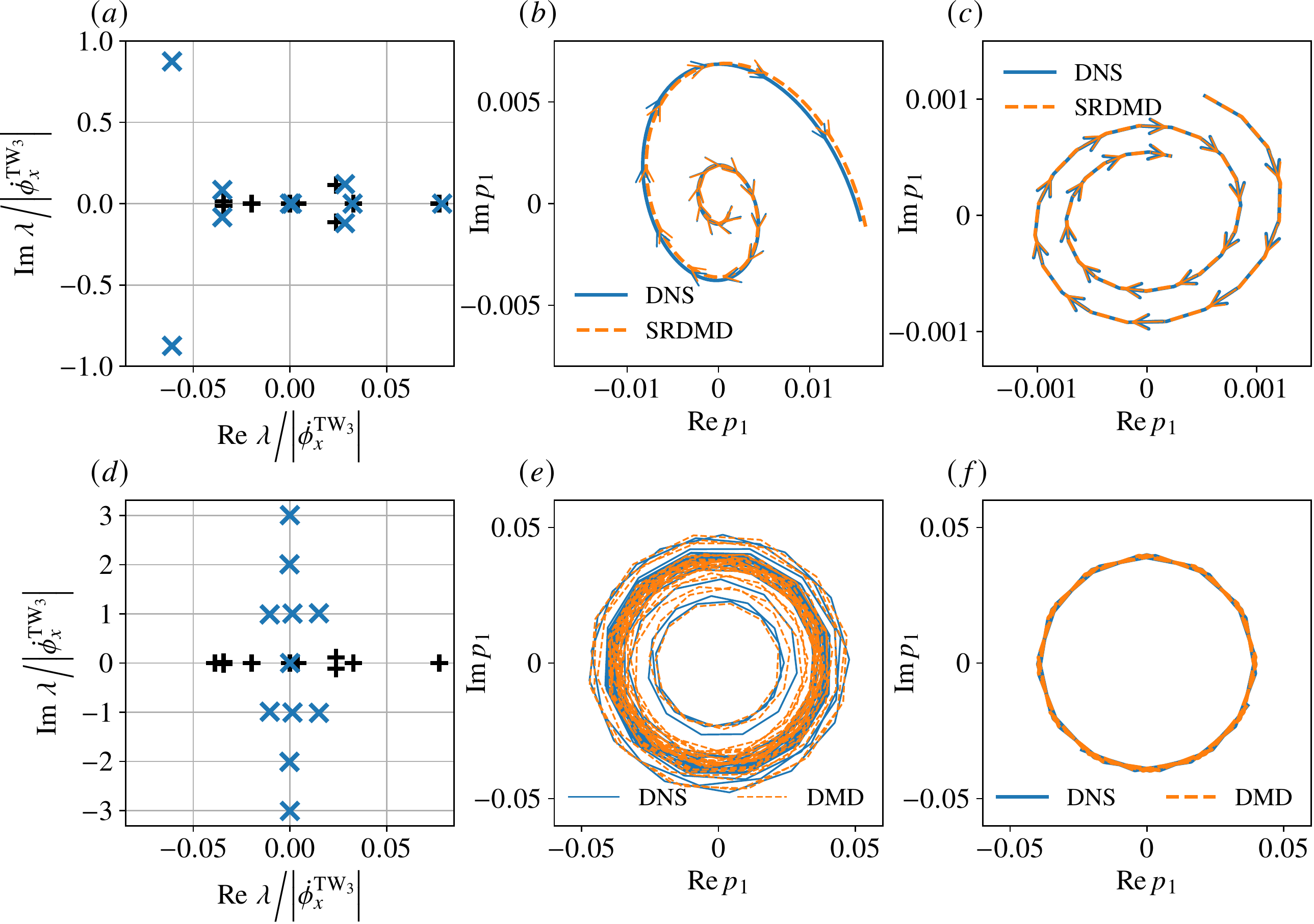}{
    (a)
    Linear stability eigenvalues (+) of the travelling wave
    $\TW_3$ approximated via Arnoldi 
    iteration and the SRDMD  exponents
    ($\times$, blue) computed from randomly perturbed 
    trajectories in $\TW_3$'s vicinity. 
    (b) 
    A spiral-out trajectory (see main text) on $\TW_3$'s 
    unstable manifold and its SRDMD approximation 
    visualized as a projection onto the leading SRDMD modes centred 
    about the marginal one.
    (c)
    A spiral-in trajectory (see main text) on $\TW_3$'s 
    stable manifold and its SRDMD approximation 
    visualized as a projection onto the leading SRDMD modes centred 
    about the marginal one.
     (d--f) same as (a--c) without symmetry
    reduction.
    Both (a) and (d) have their axes normalized by 
    $\left|\dot{\slphase}_x^{\TW_3}\right|=0.53$, the streamwise slice velocity
    of $\TW_3$, which is constant for a travelling wave.
}

As our second application, we adapt the DMD-based periodic orbit search method
of \cite{page2020searching} to
 relative periodic orbits. To this end, we simulate 
the plane-Couette flow in the HKW cell for a time interval $[0, 2000]$ and
sample the trajectory at steps of
$\ts= 0.1$.
We then slide a temporal window of fixed duration $T_w = 60$ 
in steps of $\Delta_w=5$ along the time series and compute the SRDMD
of each window using $m=100$ randomly-chosen 
snapshot pairs
with separation time $\delta t=1$.
Next, we calculate the periodicity indicator 
\citep{page2020searching}
\eq{periodicity_indicator}{
    \varepsilon (n_h) \coloneqq 
      \frac{1}{n_h \omega_f^2} 
      \sum_{j = 1}^{n_h} |\Im \dmdeval_j - j \omega_f|^2, 
}
where 
\eq{frequency}{
    \omega_f (n_h) \coloneqq \frac{2}{n_h(n_h+1)} \sum_{j}^{n_h} \Im \dmdeval_j  
} 
and the sums are carried over the $n_h$ SRDMD exponents with 
$\Re \dmdeval_j < \mu^{max}$ and 
$0 < \Im \dmdeval_1 < \ldots < \Im \dmdeval_{n_{h-1}} < \Im \dmdeval_{n_h}$.
For an exactly periodic signal, 
$\varepsilon = 0$, and the DMD exponents are purely imaginary, i.e.\ $\Re \dmdeval_j = 0$ 
\citep{rowley2009spectral}.
Thus, $\varepsilon < \varepsilon^{th}$ 
for a set of DMD exponents with real parts below a chosen threshold 
$\mu^{max}$ indicates approximate periodicity \citep{page2020searching}.
We set $n_h = 2$, $\mu^{max} = 0.1$ 
and select guesses
for relative periodic 
orbits from episodes with $\varepsilon < 10^{-4}$.
Note that the chosen subset of SRDMD exponents
contains at least one real mode and for all the flagged episodes
analysed here, we have $N_d=5$ (one real mode plus 2 complex conjugate pairs).
We experimented with higher number of harmonics, i.e. $n_h=3$ and $4$
(corresponding to $N_d=7$ and $9$), 
but we found that increasing the number of harmonics
did not provide any additional initial guess that
converged to a relative periodic orbit.
For the flagged episodes that we obtained, we use
the state
$\xi^{(g)}$
at the time instant within the window corresponding to the minimum of 
the reconstruction error, the period
$T^{(g)} = 2 \pi / \om_f$,
and the shifts
\(
    \De x^{(g)} 
    = L_{x} [ \slphase_{x}(t^{(g)} + T^{(g)}) 
             - \slphase_{x}(t^{(g)}) ] / (2 \pi)
\)
and 
\(
    \De z^{(g)} 
    = L_{z} [ \slphase_{z}(t^{(g)} + T^{(g)}) 
             - \slphase_{z}(t^{(g)}) ] / (2 \pi)
\)
as initial guesses to initiate Newton--Krylov-hookstep
\citep{viswanath2007recurrent} searches for relative 
periodic orbits.
Here, the superscript $^{(g)}$ stands for ``guess''.
 Among 16 such searches, 2 converged to the 
time-periodic solutions of the 
plane-Couette flow in the HKW cell.
One of these orbits was a known periodic orbit of plane-Couette 
flow with $(T_\mathrm{po}, \De x_\mathrm{po}, 
\De z_\mathrm{po}) 
= (64.9, 0, 0)$ and can be found 
in the \texttt{channelflow.org} database. Since our goal 
here is to illustrate the utility of symmetry reduction for dynamics with 
spatial drifts, we do not report this orbit here and turn our attention 
to the other with a nonzero drift in the streamwise direction.
\Figref{rpo79}(a,b) shows 
the SRDMD exponents and spectra of the DNS 
window which converged to a relative 
periodic orbit $\RPO_{79.4}$ with 
$(T_\mathrm{rpo}, \De x_\mathrm{rpo}, 
\De z_\mathrm{rpo}) = (79.4, 0.356 L_x, 0)$. 
Note that the converged orbit's period is approximately $4/3$s of the DMD
time window, demonstrating that SRDMD is capable of producing an initial 
condition for a relative periodic orbit search even when the orbit is not 
followed by the chaotic dynamics for a full period. 
This can also be seen by comparing the state space projections of the 
guess episode shown in \figref{rpo79}(c) to that of the converged orbit 
in \figref{rpo79}(f) onto the subspaces spanned by the leading non-marginal 
SRDMD modes.
Hence, SRDMD successfully extends \cite{page2020searching}'s periodic orbit 
detection method to the relative periodic orbits with spatial drifts.
The SRDMD of the converged orbit 
computed using snapshots separated by $\delta t = 1$ along one full period
approximates a Fourier expansion as indicated 
by the fact that the near-neutral
(with $\Re \lambda < 10^{-4}$)
exponents are
located at the harmonics 
of $f = 1 / T_\mathrm{rpo}$ as shown in \figref{rpo79}(e). This is expected since 
continuous symmetry reduction transforms the relative periodic orbit to a 
periodic one, for which the DMD
corresponds to a Fourier expansion 
\citep{rowley2009spectral}. For comparison in \figref{rpo79}(g,h), we show 
the DMD exponents corresponding to the same dataset and spectra of the relative 
periodic orbit without reducing its symmetry, which shows no resemblance to that of a 
Fourier series.
Finally, \figref{rpo79}(i) shows the state space projection of 
the converged relative periodic orbit without symmetry reduction where the separation 
of the initial and final states is clearly seen. 
Although \figref{rpo79}(i) suggests that DMD without the symmetry reduction 
still yields a decent approximation to the periodic orbit, the 
corresponding DMD spectrum has a leading
non-marginal DMD eigenvalue 
that is negative real, i.e.\ non-oscillatory, as shown by the orange marker 
in \figref{rpo79} (g) and (h). The difference of the initial and final states 
of the orbit is
also visible in Supplementary Movie 3, where the flow structures of the
initial and final states appear at the same spots only after
symmetry reduction.

\Figref{singularity}(a) shows the streamwise slice phase velocity 
$\dot{\slphase}_x$ along one period of $\RPO_{79.4}$ approximated by 
finite differences 
$\dot{\slphase}_x \approx 
(\slphase_x (t + \sldt) - \slphase_x (t))/\sldt$ 
and the reconstruction equation \eqref{slicephasex}, for which we approximated 
the state space velocity as 
$\partial_t \vu \approx (\vu (t +  \sldt) - \vu (t)) /  \sldt$
and used \(\sldt=0.01\) for both finite difference approximations. 
As shown, near the midpoint of the shown time window, the phase velocity 
momentarily approaches $\dot{\phi} \approx -10$, the effect of which can 
also be seen as a fast drift in Supplementary Movie 3. 
\Figref{singularity}(b) shows that this instance coincides with a 
near-zero of the reconstruction equation's
\eqref{slicephasex}
denominator, which accentuates the instantaneous decrease of the 
numerator term as shown in \figref{singularity}(c). This illustrates an 
important property of our symmetry reduction method: even though 
instantaneous fast oscillations can be introduced as an artefact, the net 
drift of an invariant solution, in this case a relative periodic orbit, is
zero for the totality of the orbit, i.e.\ one full period.

\fig{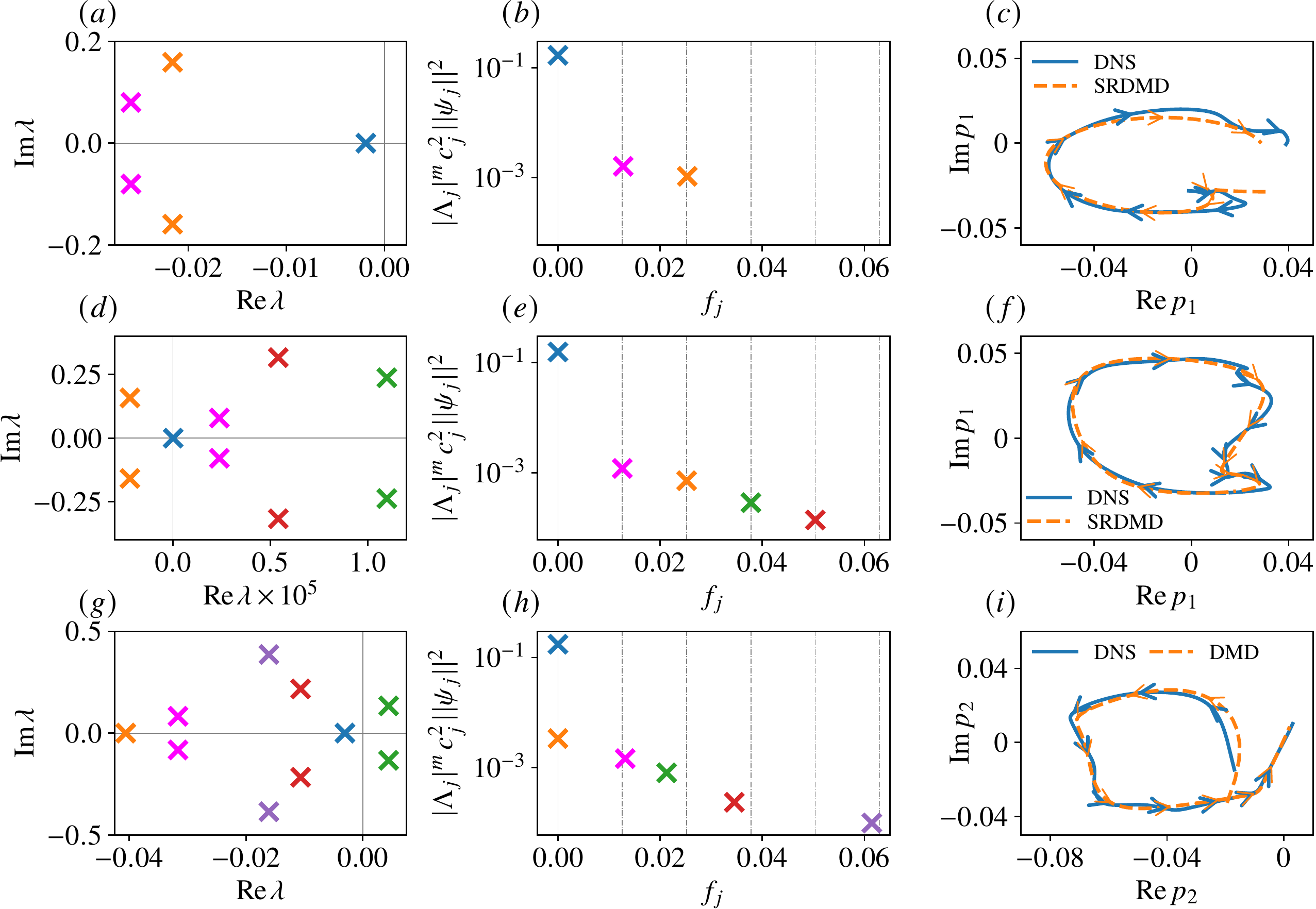}{
    (a) SRDMD exponents and (b) spectrum of the episode from 
    which the initial guess for a relative periodic orbit $\RPO_{79.4}$
    is constructed. 
    (c) State space projection of this episode and its SRDMD approximation 
    onto the plane spanned by the leading non-marginal SRDMD modes centred around 
    the marginal mode.
    (d) SRDMD exponents and (e) spectrum of the $\RPO_{79.4}$.
    (f) State space projection of the $\RPO_{79.4}$ and its SRDMD 
    approximation.  
    (g) DMD exponents, (h) DMD spectrum of the same orbit
    without symmetry reduction; and (i) the corresponding state space projections. 
    In (b), (e) and (h) $f_j = |\Im \lambda_j| / 2\pi$ and the dashed vertical lines 
    correspond to multiples of the RPO's fundamental frequency 
    $2\pi / T_\mathrm{rpo}$, where $T_\mathrm{rpo}$ is the period of
    $\RPO_{79.4}$.
}

\fig{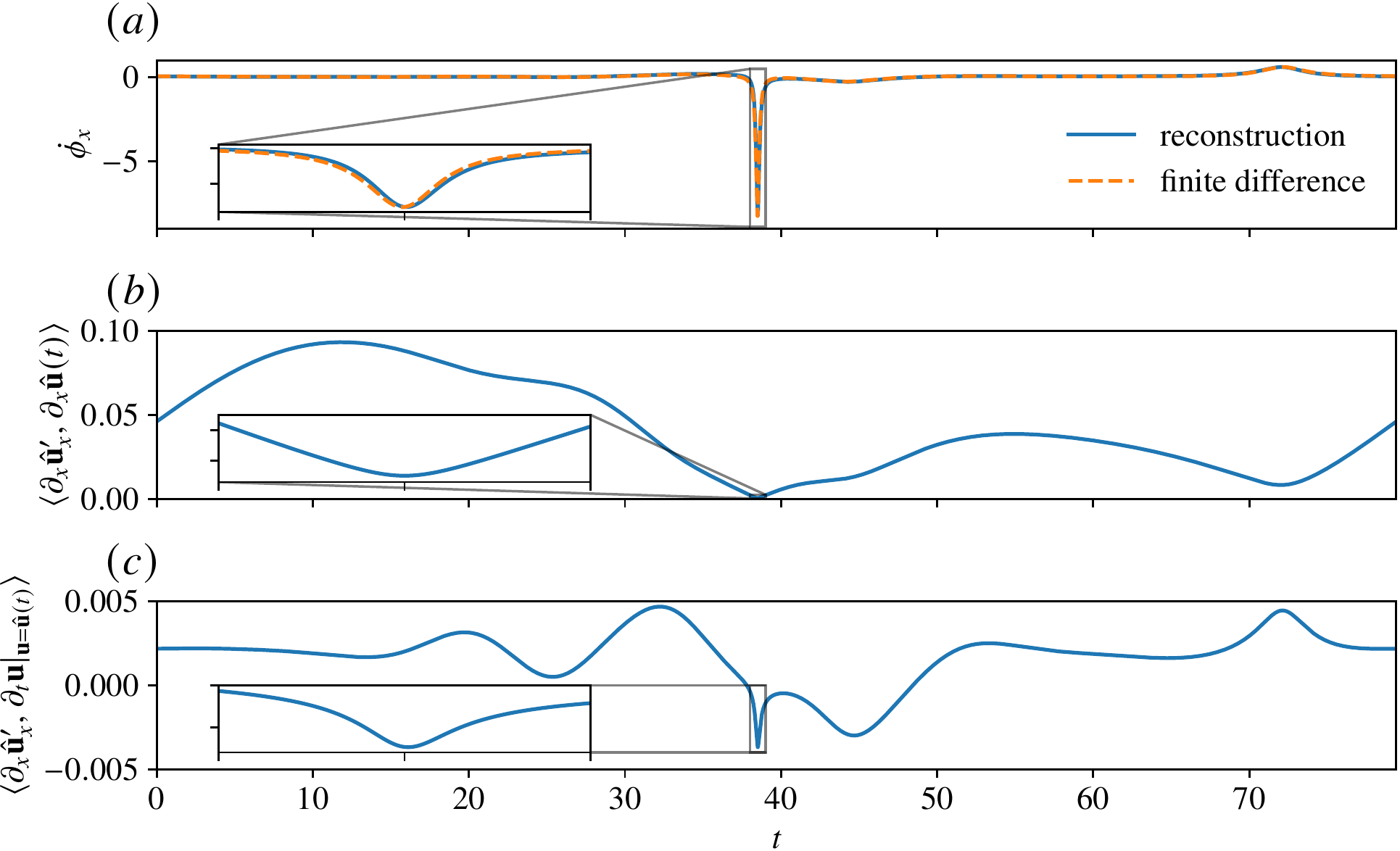}{
        Time series from a full period of $\RPO_{79.4}$.
        (a) The streamwise slice velocity found from the reconstruction equation
        \eqref{slicephasex} vs.\ its approximation via finite differences of 
        slice phases.
        (b) Denominator and (c) numerator of the reconstruction 
        equation \eqref{slicephasex}.
}

\section{Locally linear approximations by SRDMD}
\label{sec/linear}

Through the examples of the previous section, we demonstrated how symmetry reduction
enables DMD to capture the transitional/low-$Re$ dynamics in the vicinity of the
relative invariant solutions of plane-Couette flow with spatial drifts. We now
turn our attention to plane-Poiseuille flow domains P2K and P5K, corresponding
to $\Rey = 2000$
and $\Rey = 5000$, respectively, see \tabref{setup}.
Both of these domains are
significantly
more complex than the settings considered in the previous section, 
hence, searching for invariant solutions in them is impractical.
As we shall illustrate, one can nevertheless utilize SRDMD in this problem to
discover turbulent episodes that can be transiently approximated by a
low-dimensional linear expansion. 

We consider turbulent channel flow
simulations in P2K and P5K, each spanning a time interval 
$[0, 2000]$. We
compute SRDMD of the data
sets 
    \(
        \Xi  = \left[ \xi_0, \xi_1, \ldots, \xi_{m-1} \right] \mbox{ and }
        \Xi' = \left[ \xi_1, \xi_2, \ldots, \xi_{m} \right]
    \)
where $\xi_n$ are the symmetry-reduced fluid states 
sampled at $\delta t$ over sliding time windows of length 
$T_w$.
In P2K, we took $\de t = 1$ and $T_w \in \{30, 60, 100\}$, whereas 
in P5K, we set these to half of their values, i.e.\
$\de t = 0.5$ and $T_w \in \{15, 30, 50\}$, approximately matching the P2K 
values in wall units.
In order to compare different
episodes and window lengths, we construct SRDMD approximations \eqref{dmdApprox}
with $\Ndmd = 11$ (or 12, depending on the number of complex conjugate
exponents in the dominant part of the 
SRDMD spectrum, such that if a complex mode is in the 
spectrum, so is its complex conjugate). We evaluate the accuracy 
of SRDMD by measuring the residual
\eq{dmdRes}{
  \dmdRes (t) = \frac{1}{m} \sum_{k = 0}^{m -
      1} \frac{\| \dmdApprox(t + k\, \delta t)
      -\xi(t + k\, \delta t) \|}{ \| 
      \xi(t + k\, \delta t) \| }\, , 
      }
which is the mean relative error of the SRDMD approximation 
\eqref{dmdApprox}
to the symmetry-reduced DNS states in the same
time window $[t, t+T_w)$. 
\Figref{residual} shows the residuals \eqref{dmdRes} of our SRDMD 
approximations with $N_d \le 12$ to the sliding windows of the turbulent 
channel flow data.
Low-error episodes are detected along the turbulent trajectory and
appear to be clustered around certain time instants, for example
around $t \approx 250$, 950 and 1250 in \figref{residual}(a),
thus signalling portions of the turbulent evolution
that can be well captured by a reduced linear expansion.  
As expected, a linear approximation is more successful for shorter times, and
accordingly the dips in the $\overline{\mathcal{R}}(t)$ curves are most marked
for the shortest window lengths $T_w=30$ 
(P2K) and $T_w = 15$ (P5K), although they are still
distinguishable for longer time windows.
In the following,
 we decided to focus on  
$T_w = 60$ and $T_w = 30$ in P2K and P5K, respectively
(orange curves in \figref{residual}),
which are the longest time windows 
of those analysed above
where clear low-error episodes ($\dmdRes < 0.1$) were detected.
In the rest of this section, 
different dynamical behaviours captured by 
the SRDMD approximations at these time window lengths 
will be illustrated.

\fig{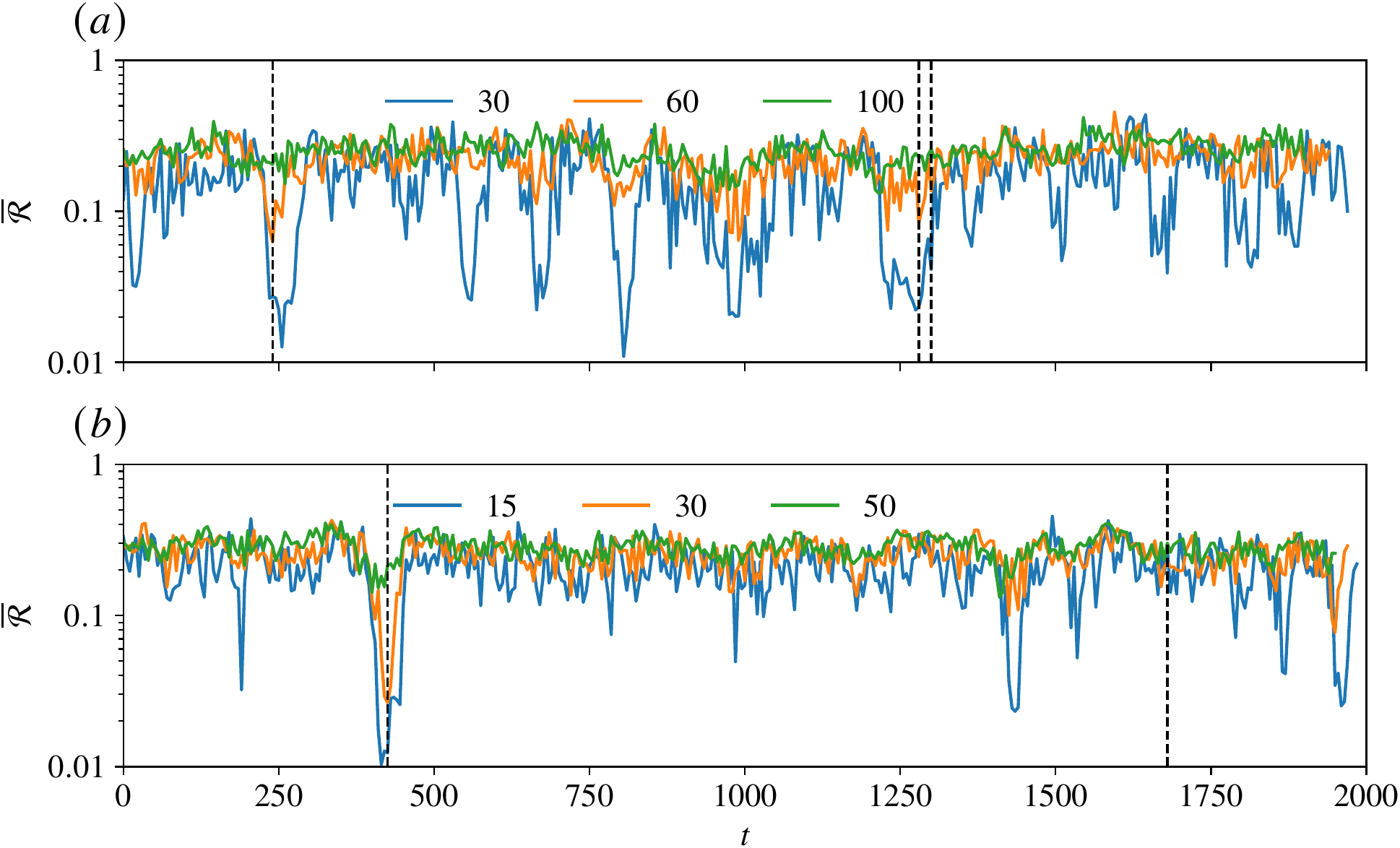}{
    SRDMD residuals \eqref{dmdRes} in
    (a) the P2K domain for time windows of duration $T_w=30$, 60, and 100 and 
    (b) the P5K domain for $T_w=15$, 30, 50. Time windows were slided across 
    the time series in steps of $\Delta_w=5$. 
        Dashed vertical lines correspond to 
        the initial times of the
        windows that are presented in
        figures (\ref{fig/spectrum}--\ref{fig/snapshots-5k}).
}

\Figref{spectrum}(a) shows the SRDMD spectrum of the data window
corresponding to $t \in [1280, 1340)$ in P2K,
where the coloured crosses indicate the 
dominant part of the spectrum and the black ones show the first three discarded modes. 
To illustrate the flow structures captured by SRDMD, \figref{spectrum}(b--d) shows
three-dimensional visualizations of 
$\dmdMode_0$, $\Re \dmdMode_1$ and $\Im \dmdMode_1$, respectively. In 
\figref{spectrum}(b--d) and the rest of the flow
visualizations of this paper, the red/blue isosurfaces show $u = 0.5\, \max /
\min u$ and the green/purple isosurfaces show $\omega_x = 0.5 \, \max / \min
\omega_x $, where $\omega_x$ is the streamwise vorticity. Once again, for comparison 
in \figref{spectrum}(e) we show the DMD spectrum, without 
symmetry reduction, of the same episode. For this computation, we needed 
a temporal resolution of
$\delta t = 0.1$ because the timescale associated with advection is much faster
than those of the coherent structures.
As visualized in \figref{spectrum} (b--d) the SRDMD modes
represent the full complexity of the turbulent episode that they approximate
and, remarkably, their \emph{linear} time evolution exhibit generation and
disappearance of coherent structures as shown in Supplementary Movie 4.
As shown in \figref{spectrum}(e) in the absence of symmetry reduction the dominant 
frequency of the spectrum appears near the drift frequency 
$f_{d} = U_b/L_x \approx 0.23$, where $U_b=2/3$ is the bulk velocity 
marked by a dashed line in \figref{spectrum} (e).
In contrast to the complex flow structures captured by the SRDMD, in this case the marginal 
mode (\figref{spectrum}(f)) shows elongated structures and the leading non-marginal DMD 
mode aligns with the first streamwise Fourier mode (\figref{spectrum}(g,h)) 
as signified by the fact that  $\Im \psi_1$ 
(\figref{spectrum}(g)) is virtually the same 
as $\Re \psi_1$ (\figref{spectrum}(h)) up to a 
quarter-domain shift in the $x$ direction.
We see that even for the limited time window that we consider here, the 
streamwise drift of channel flow completely dominates its DMD.

\fig{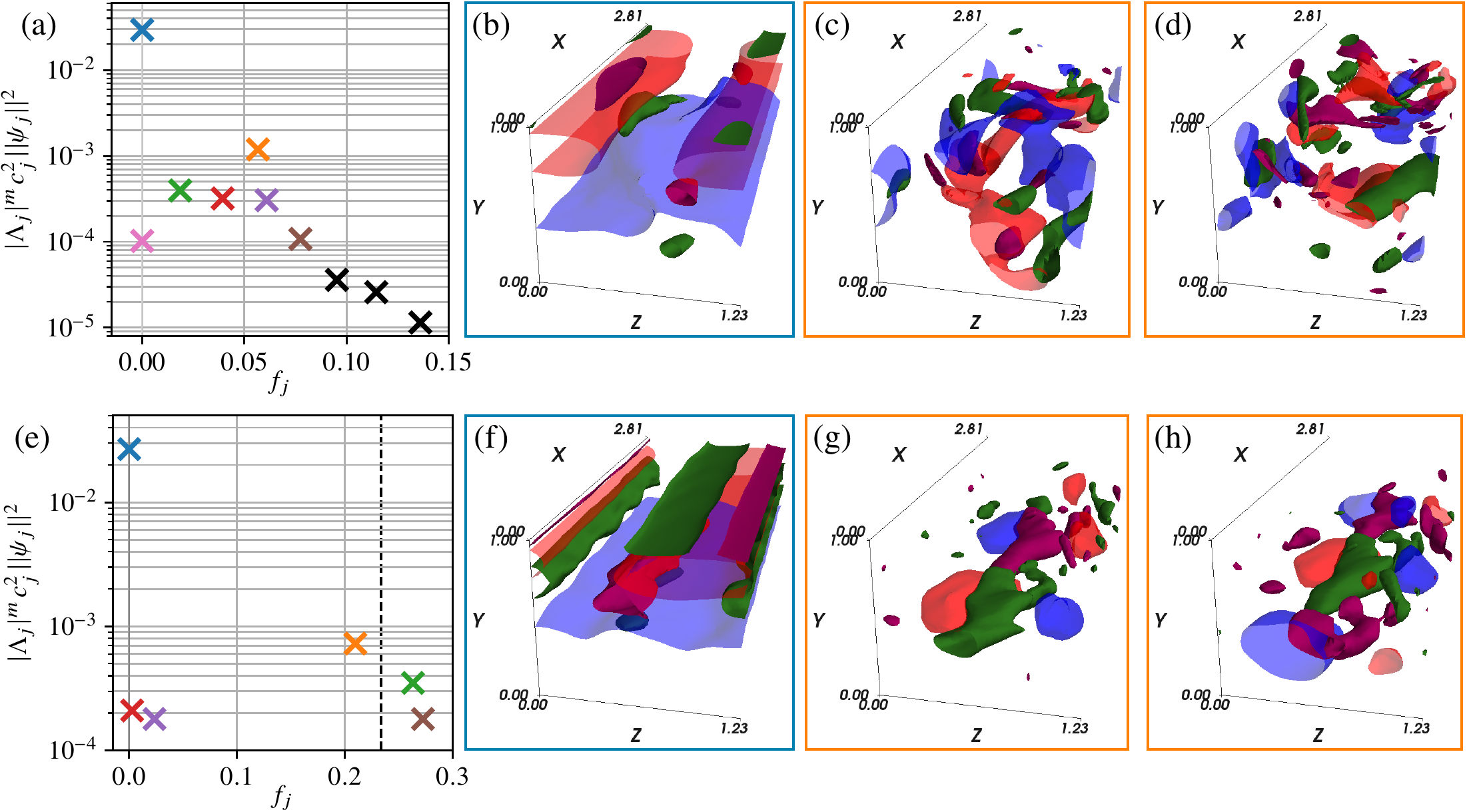}{
    SRDMD in the time window $t \in [1280,1340)$ of the P2K 
    domain.
    (a) Normalized SRDMD spectrum
    where $f_j = |\Im \lambda_j| / 2\pi$. The coloured 
    symbols correspond to the modes that are
    included in the sum \eqref{dmdApprox}, while the black
    symbols are the first three discarded modes. 
    (b--d) Three-dimensional visualizations of the SRDMD modes
    $\dmdMode_0$ (b), $\Re \dmdMode_1$ (c) and 
    $\Im \dmdMode_1$ (d), where the red/blue 
    isosurfaces show $u = 0.5\, \max / \min u$ and the green/purple 
    indicate the streamwise vorticity isosurfaces 
    $\omega_x = 0.5 \, \max / \min \omega_x $.
    (e) DMD spectrum (without symmetry reduction) of the same episode.
    The dashed vertical line corresponds to the drift frequency 
    $f_{d} = U_b/L_x$, where \(U_b\) is the bulk velocity.
    (f--h) Three-dimensional visualizations of the DMD modes.
    The colours of the bounding boxes in three-dimensional 
    visualizations correspond to
    the (SR)DMD modes marked with the same 
    colour on the spectra on the left.
}

\fig{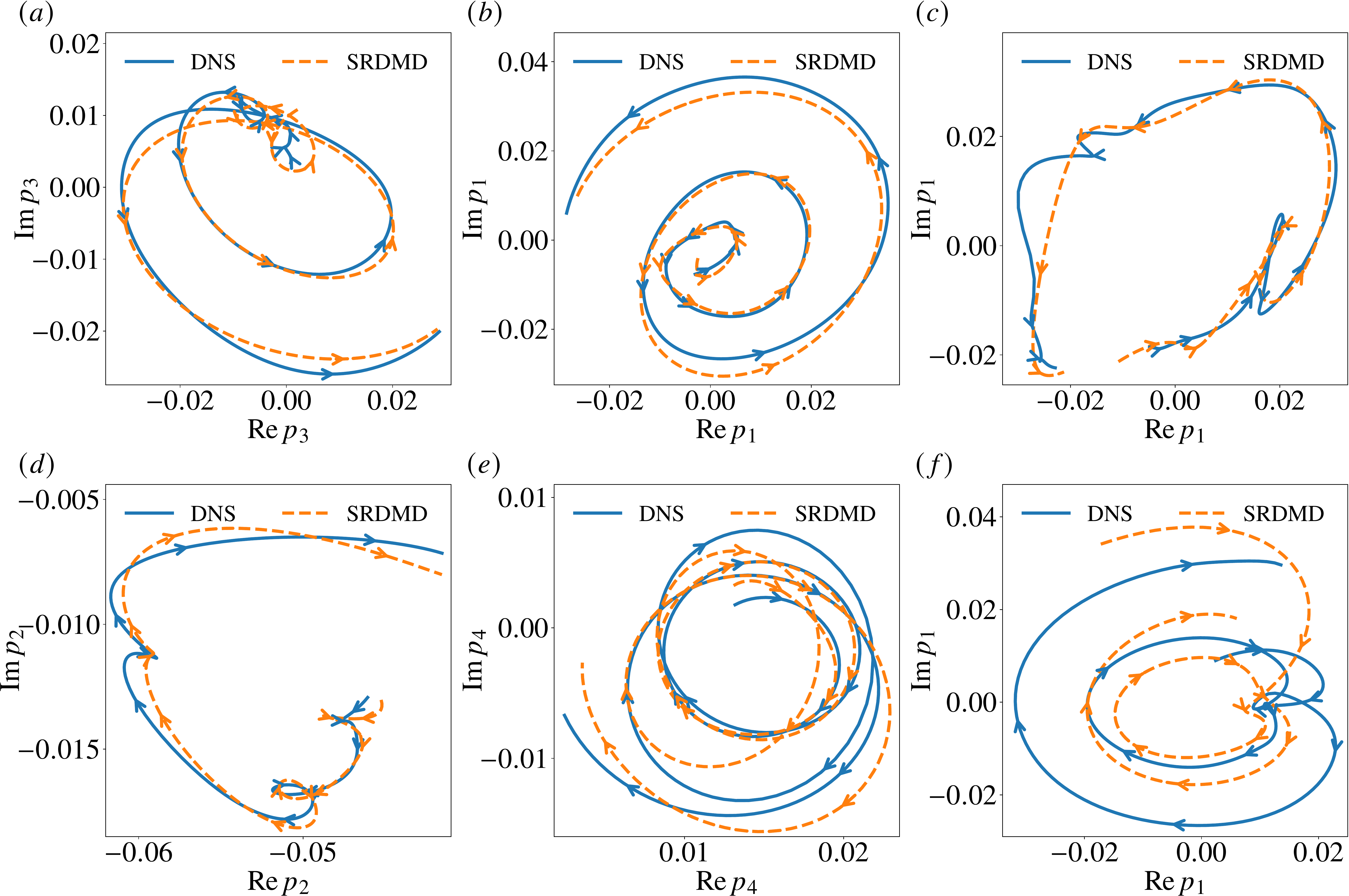}{
State space projections of DNS trajectories and their SRDMD 
approximations from (a--c) the P2K and (d--f) 
the P5K domains
onto complex SRDMD modes centred
around the marginal modes. 
The episodes correspond to 
(a) $t \in [240, 300)$, (b) $t \in [1280, 1340)$, (c) $t\in[1300,1360)$
in P2K and 
(d,e) $t \in [425, 455)$, and (f) $t\in[1680,1710)$ in P5K. 
}

As illustrations of different dynamical regimes captured by SRDMD, in
\figref{projections} we show state space projections of 
several time windows and their SRDMD 
approximations onto the subspaces
spanned by complex conjugate SRDMD modes. In \figref{projections}(a,b), we 
see spiral-out/in dynamics corresponding to the low-error episodes in 
\figref{residual}(a), reminiscent of 
similar episodes that we illustrated for a travelling wave in \figref{TW3} 
(c,d). In \figref{projections}(c), we show a nearly-periodic trajectory 
at $\Rey = 2000$, which we detected at a minimum of the periodicity indicator 
\eqref{periodicity_indicator} with $n_h=4$,
$\mu^{max} = 0.1$ and $\varepsilon^{th} = 2 \times 10^{-3}$.
As expected, at $\Rey = 5000$ the dynamics is significantly more complex, 
nevertheless the resemblance of state space projections of DNS data and their 
SRDMD approximations can also be seen in \figref{projections}(a--c). 
Differently from \figref{projections}(a,b), here, we do not see simple 
spiral-in/out dynamics, but several instabilities at play, as illustrated 
by \figref{projections}(d,e) which shows the projections of the same low-error
episode into the subspaces spanned by the corresponding SRDMD modes 
$\ps_2$ and $\ps_4$. Finally, \figref{projections}(f) shows a nearly-periodic 
episode in the P5K domain detected at a minimum of the periodicity indicator 
\eqref{periodicity_indicator} using the same parameters as P2K.

For the spiral-out event starting at $t=1280$ in P2K (\figref{projections}(b)), we
compare the evolution of the flow structures reconstructed using SRDMD to those
of the original turbulent dynamics, see \figref{snapshots} and Supplementary
Movie 4. The SRDMD spectrum and the first two SRDMD 
modes for this time window were displayed in \figref{spectrum}. 
As shown in \figref{snapshots},  
SRDMD can capture the evolution of streaks and rolls,
visualized as isosurfaces of streamwise velocity and streamwise
vorticity, respectively, with only 12 modes. In particular, it can
capture the initial growth of the rolls which then break up into
smaller structures and appear to decay towards the end of the time
window while the streaks start meandering. 

\fig{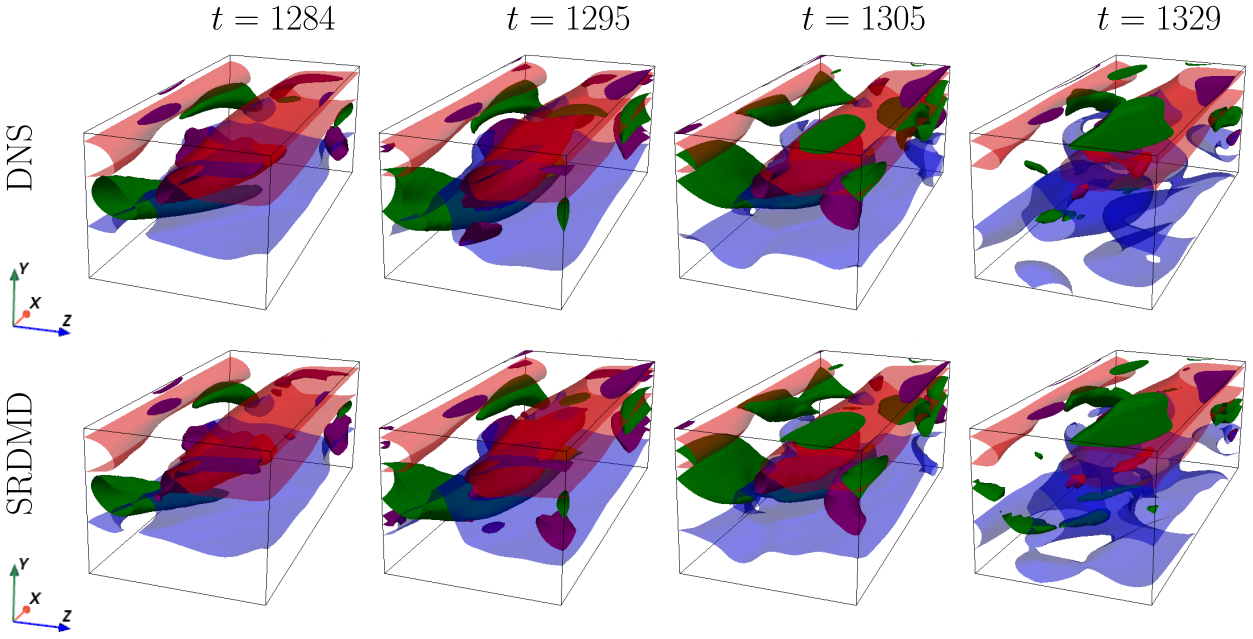}{
    Three-dimensional visualizations of symmetry-reduced flow 
    states and their SRDMD approximations for the time window 
    $t \in [1280, 1340)$ in the P2K
    domain.  \label{snapshots}
}

As our final illustration, in \figref{snapshots-5k}, we show flow 
states and their SRDMD approximations corresponding to the time 
window $[425, 455)$ of the P5K domain. Although the one-to-one 
correspondence of vortical structures
is not as evident as in the P2K configuration, 
the streak evolution is still 
captured by the SRDMD approximation.

\fig{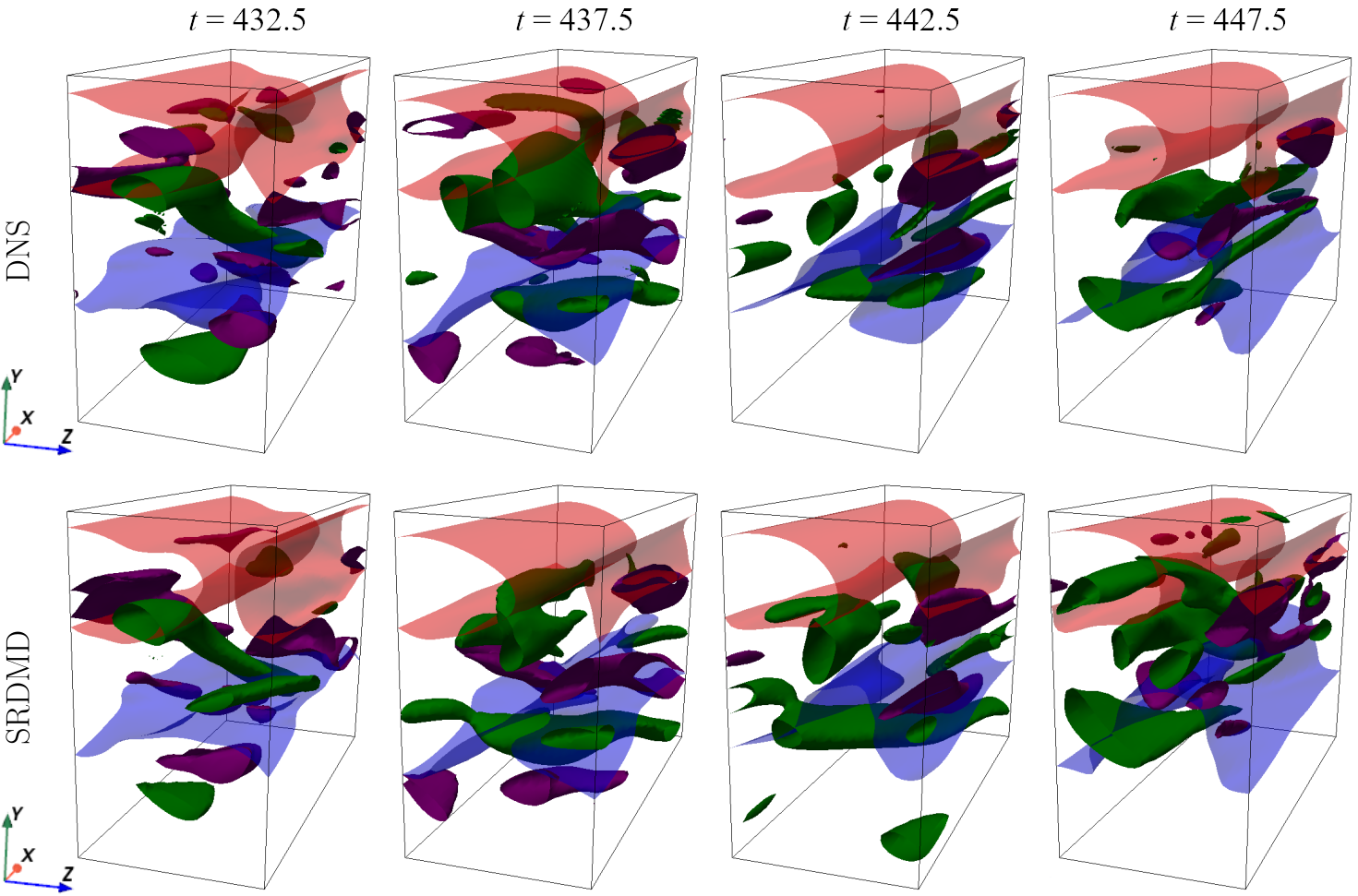}{
    Three-dimensional visualizations of symmetry-reduced flow 
    states and their SRDMD approximations for the time window 
    $t \in [425, 455)$ in the P5K
    domain.  \label{snapshots-5k}
}

\section{Conclusion and outlook}
\label{sec/conclusions}

In this paper, we developed a continuous symmetry reduction method for
three-dimensional flows in rectangular channels and applied it to
plane-Couette and -Poiseuille simulations at various \Rey\ to illustrate the
necessity of symmetry reduction for preprocessing the data prior to DMD. 
We showed in \secref{relinv} that the combination of symmetry reduction 
and DMD, which we named SRDMD, yields linear modal expansions that capture 
the stable/unstable dynamics in the vicinity of a travelling wave whereas 
the standard DMD of the same data was dominated by the drifts. Moreover, 
utilizing SRDMD, we extended \cite{page2020searching}'s DMD-based periodic 
orbit search method to relative periodic orbits and showed that a guess 
for a relative periodic orbit can be generated by SRDMD even when 
turbulence does not follow the relative periodic orbit for a full period, 
enabling relative periodic orbit searches that would have not been 
possible via recurrent flow analysis \citep{chandler2013invariant}. 
In the light of recent 
evidence \citep{yalniz2021coarse,krygier2021exact,crowley2022turbulence} 
that the approaches of turbulence to periodic solutions are often not 
for a full period, we think that SRDMD-based approaches can 
accelerate 
relative periodic orbit searches in future studies.

One of the main motivations behind studying the invariant solutions of
the Navier--Stokes equations follows from the observation that, together with
their stable and unstable manifolds, the invariant solutions provide
``efficient'' bases for approximating chaotic dynamics in their vicinity.
This is also illustrated in \figref{rpo79} of the present paper where the
state space projection of the turbulent trajectory in \figref{rpo79}(c)
shows the features of the nearby relative periodic orbit, shown in
\figref{rpo79}(f). The main message that we hope to convey with the present
paper is that such seemingly simple dynamics can also be found in more
complicated settings, such as the turbulent plane-Poiseuille flow that we
studied here. 
We would like to point out the similarities 
between the Poiseuille flow state space projections
shown in \figref{projections}(a,b) and those in the vicinity of a simple 
travelling wave shown in \figref{TW3}(b,c). Although the spatial 
structures captured by the SRDMD modes are considerably complex as 
illustrated in \figref{spectrum}(b--c) and \figref{snapshots} in the Poiseuille 
case, the temporal dynamics is surprisingly similar to that in the vicinity of 
a travelling wave. Similarly, the state space projection of the 
nearly-periodic Poiseuille flow episode in \figref{projections}(c) 
is reminiscent of that of an actual shadowing of a relative periodic orbit by 
turbulence shown in \figref{rpo79}(c). 
In the light of these observations, we speculate that
a modelling paradigm wherein one searches for simple data-driven models that
apply to specific state space regions could be a viable strategy for extending
dynamical systems approaches to turbulence to regimes that are more complicated
than those in which invariant solutions can be found.

While symmetry reduction enables us to apply DMD to flows with
continuous symmetries, it, of course, does not 
eliminate limitations that are
intrinsic to DMD itself. One of these is already apparent in
\figref{residual} where the episodes with low 
reconstruction error mostly
disappear when the SRDMD window is extended to $T_w = 100$ 
and $T_w = 50$ in the P2K and P5K domains, respectively. In general, we
do not expect to find low-dimensional SRDMD approximations for longer time
horizons since, even though DMD modes are
non-orthogonal
and can represent nonlinear processes such as generation of vortices,
the temporal dynamics constructed by them is still linear.
Recently, \cite{linot2020deep} has shown that a combination of symmetry
reduction, an autoencoder and a reservoir computer can be used to 
approximate the spatiotemporally chaotic dynamics of the one-dimensional 
Kuramoto--Sivashinsky equation. Replacing DMD with such deep learning methods 
can extend the prediction horizon beyond what can be approximated by SRDMD. 
Similar to the finite-time horizon, we also expect SRDMD to perform poorly 
in larger domains since turbulent flows have finite correlation lengths. 
This can be readily seen comparing the snapshots in 
\figref{snapshots} to those in \figref{snapshots-5k} where the height 
difference of the two domains, which have the same length and width in wall 
units, is apparent. Consequently, more flow structures are present in the 
wall-normal extent of the $\Rey=5000$ domain and the corresponding 
SRDMD approximations in \figref{projections}(b,e) are considerably more 
complex. One obvious remedy to this might be replacing the $L_2$ 
inner product \eqref{norm} which we used here with one incorporating 
a spatial filter that isolates a shorter wall-normal extent, or any 
other dynamical region of interest if SRDMD is applied to a flow 
in a larger domain.

Finally, we would like to note that in addition to being the first
continuous symmetry reduction method for channel flows, a novel aspect of the
present formulation is our optimization-based determination of the wall-normal
dependence of the slice templates. Previous implementations
\citep{willis2016symmetry,budanur2017heteroclinic,budanur2018complexity} of the
first Fourier mode slice in three-dimensional pipe flows in which both
homogeneous (axial and azimuthal) and inhomogeneous (radial) directions are
present relied on trial-and-error in determining the radial dependence of slice
templates. The template optimization procedure developed here provides a
systematic alternative to this which we recommend for future applications of the
first Fourier mode slice in similar setups. In hopes to lower the technical 
barriers for such future applications, we provide our codes in 
\cite{yalniz2022srdmd} as examples.

\backsection[Supplementary data]
    {Supplementary movies are available at [URL will be inserted by the publisher.]}
\backsection[Funding]
    {E.\ Marensi acknowledges funding from the ISTplus fellowship programme.
    G.\ Yaln\i z and B.\ Hof acknowledge a grant from the Simons Foundation 
    (662960, BH).}
\backsection[Declaration of interests]{The authors report no conflict of interest.}
\backsection[Author ORCID]{
    E.\ Marensi, \url{https://orcid.org/0000-0001-7173-4923};
    G.\ Yaln\i z, \url{https://orcid.org/0000-0002-8490-9312}; 
    B.\ Hof, \url{https://orcid.org/0000-0003-2057-2754}; 
    N.\ B.\ Budanur, \url{https://orcid.org/0000-0003-0423-5010}}

\printbibliography

\end{document}